\documentclass{jaa}
\usepackage{natbib}
\usepackage{graphicx}
\usepackage{subcaption}

\begin{document}\sloppy

\title{Journey of  X-ray astronomy: 
Indian perspectives}


\author{A. R. Rao  \textsuperscript{1,*}}
\affilOne{\textsuperscript{1}Department of Astronomy and Astrophysics, Tata Institute of Fundamental Research, Mumbai 400 005, India.\\}


\twocolumn[{

\maketitle

\corres{arrao@tifr.res.in a.raghu.rao@gmail.com}

\msinfo{}{}

\begin{abstract}
X-ray astronomy is a mature area of observational astronomy. 
After the discovery of the first non-solar X-ray source in 1962, X-ray astronomy proliferated during the Apollo era's space race. Then, it matured as an established area of research during the period of Great Observatories, and now it has become an indispensable tool to understand a wide variety of astrophysical phenomena. Consequently, in recent times, niche observational areas in X-ray astronomy have been explored, and attempts have been made to expand the sensitivity of observations vastly.
  India was an active partner in the growth of X-ray astronomy. In the initial years, India leveraged its expertise in balloon technology to get significant results in the research area of hard X-ray astronomy. 
 During the rapid growth phase of X-ray astronomy, India made divergent all-round efforts. 
  Later on, however, the technical expertise available in India was insufficient to compete with the highly sophisticated satellite experiments from around the world. During this phase, work in X-ray astronomy continued in a few low-key experiments, eventually resulting in the launch of India's first multi-wavelength astronomical satellite, AstroSat, in 2015. In this article, I will trace the journey of X-ray astronomy and the developments in the Indian context.   I will also explore the sociological aspects of the growth of X-ray astronomy, and, in the end, I will present a speculative sketch of the future of X-ray astronomy with an emphasis on the Indian contribution.
  \end{abstract}


}]

\onecolumn


\doinum{12.3456/s78910-011-012-3}
\artcitid{\#\#\#\#}
\volnum{000}
\year{0000}
\pgrange{1--}
\setcounter{page}{1}
\lp{1}

\section{Introduction}

 X-ray astronomy is one of the major branches of space astronomy, astronomy that could only be done by sending instruments outside the protective blanket of the Earth's atmosphere \citep[the textbook `The X-ray Universe' ,][gives an excellent historical account of the birth and growth of X-ray astronomy]{GiacconiBook}.
 
 Even before acquiring the ability to send instruments above the Earth's atmosphere, it was surmised that the Sun emits X-rays. The optical spectral characterisation of the solar corona indicated the existence of a hot, low-density plasma as a potential emitter of X-rays. The disturbances in the ionosphere were found to be related to solar activity, and it was surmised that ionising radiations from the Sun (UV and X-rays) must be the cause for such disturbances. Hence, when the captured German V2 rockets were made available for scientific use in the USA, the first use was to measure X-rays from the Sun using rudimentary X-ray detectors like Geiger counters. There was, however,  little expectation of detecting bright X-rays from any other objects in the sky because the amount of solar  X-ray radiation was negligibly small compared to its optical emission, and an extrapolation of the then-prevailing wisdom on stellar structure did not predict any measurable X-rays from other sources in the sky.
  
The great space race of the sixties between the USA and the USSR saw the establishment of the civilian NASA program in the USA, which started the famed Apollo program. 
 A vast amount of resources were channelled into the space programs. Several groups started benefiting from this support, the prime among them being the American Science \& Engineering (AS\&E), established by  MIT as a spin-off company. Riccardo Giacconi, who headed the Space Science Division at AS\&E, 
  was a trained Cosmic Ray physicist, and he was fascinated by the abundant possibilities of space technology for scientific use. He improved the X-ray detection techniques and proposed to NASA that reflected  X-rays from the moon could be detected by making a rocket flight. This proposal was turned down, ostensibly due to the dim possibility of detecting any X-rays from the moon. Giacconi, however, re-proposed the same experiment to the Air Force, which was approved. Though no X-rays were detected from the moon, a very bright extra-solar X-ray source, Sco X-1, was discovered. This heralded the birth of X-ray astronomy. Giacconi made further ambitious plans for X-ray astronomy and was rewarded with a Nobel Prize in Physics in 2002 `for pioneering contributions to astrophysics, which have led to the discovery of cosmic X-ray sources'.

\subsection{X-ray astronomy: new trends in doing science}
 
The birth of  X-ray astronomy, however, marked the beginning of two critical trends in scientific research. 

The twentieth century saw an explosion of our knowledge of the world around us. 
Tremendous developments and insights were obtained in fundamental physics, including relativity and quantum mechanics. These were duly applied to understand the world around us, including the cosmos.   Large optical telescopes routinely collected high-quality spectral data, which were understood from fundamental Physics, leading to very robust concepts of how stars formed, evolved and radiated. Similarly, the rigorous application of the physical principles to particle physics brought in the Standard Model of particle physics. Thus, there was some element of confidence, or complacency, in the way new phenomena are addressed: with a feeling that any new observation must surely be able to be explained by the known laws of physics.

 X-ray astronomy brought unexplained and new phenomena to the fore. Chandrasekhar did theoretically demonstrate that White Dwarfs, the degenerate state of the stars, will have their radius decreasing with their mass and will be zero for a limiting mass called the Chandrasekar mass limit, thus hinting at some state of zero size and large mass. Applying Hubble's law showed objects with huge luminosity and small sizes, known as Quasars. However, X-ray astronomy made the romantic discovery of some unseen objects, more massive than any degenerate star, emitting copious amounts of X-rays. The discovery of such X-ray binaries nudged the community to accept the notion of `back holes'. Detection of mysterious bright objects called the `gamma-ray bursts', possibly coming from cosmological distances, added to the general aura of seeing something mysterious in the cosmos.
  
Another aspect brought about by X-ray astronomy is the full use of corporate culture in doing science.

The extreme difficulties of sending something to space and making it work are generally solved with the full use of planned and objective work with attention to detail for every aspect of the work. Objectively looking at the task at hand to optimize resources, time,  and product quality can be dubbed as 
 extreme corporatization of scientific activities. X-ray astronomy hugely benefited from this method by sending instruments of ever-increasing complexity and sophistication to space and evolving the practice of optimally using the data by disseminating it to a vast astrophysical community.
 
 The dichotomy of corporate science and the sense of complacency in understanding new phenomena by known theories is quite prevalent in X-ray astronomy and perhaps in other branches of scientific research. I had pointed out this earlier while reviewing a volume of Annual Review of Astronomy and Astrophysics \citep{RaoCurrentSci}.
  Some trends in astronomy were noted: corporatization of science and commodification of science products. One of the positive aspects of this method is the ability to quickly realize well-defined problems (like measuring the afterglows of Gamma-ray bursts by a quick-moving satellite called  Swift or detecting gamma-ray pulsars using a detector with superior sensitivity). The negative aspect was `suppressing dissent and a lack of progress in long term and deep thoughts'.
  
\subsection{X-ray astronomy in India}
 
  When X-ray astronomy was born in 1962 due to the serendipitous discovery of an extremely bright X-ray source called Sco X-1 by Riccardo Giacconi in a legendary rocket experiment,  India was in an excellent position to participate in this exciting journey. After its independence, India allocated a good amount of resources for doing fundamental research and space research. One of the pioneering examples of fundamental research in India is the founding of the Tata Institute of Fundamental Research (TIFR).
 
Tata Institute of Fundamental Research  
(TIFR)  was founded by the visionary Indian physicist Homi Bhabha in 1945, two years before India gained independence. He wanted to establish `a vigorous school of research in fundamental physics' and grow it into `a school of physics comparable with the best anywhere'. He could convince the political establishment of the newly born relatively poor country that investing in the study of `abstract subjects like advanced physics, mathematics, and astrophysics' is worthwhile because practical advances of a society in the modern times `have their origin in fundamental research'. TIFR was liberally funded, and before his untimely death in 1966, Homi Bhabha could grow several areas of research, like mathematics, theoretical physics, and experimental cosmic-ray studies into maturity and also venture into new areas like computer science, biological science, and radio astronomy. Identifying talented youngsters from across the county, rigorously training them, providing them total freedom to pursue their dreams, and providing them full material support was the core method by which Homi Bhabha built this research institute. This method was fully augmented and complemented by a proactive approach of scouting for established scientists from all across the globe and convincing them to join TIFR or at least have a close interaction.
 
 The spirit of nation-building and embarking on new areas was quite widespread. Vikram Sarabhai, ten years younger than Homi Bhabha, was a multifaceted, talented individual interested in industry, management, and science. He was a trained cosmic ray physicist, and in 1947, he founded the Physical Research Laboratory (PRL) in Ahmedabad, India. PRL is thought to be the cradle of space science in India. Sarabhai, apart from his various other activities like starting nuclear reactors in India (along with Homi Bhabha), was responsible for establishing the India Space Research Organisation (ISRO) in 1969 with the express purpose of making India `...second to none in the application of advanced technology to meet the problems of man and society'.
 
 Thus, there was enough expertise available in India in the sixties so that India could be a part of the exciting journey of X-ray astronomy.  
 In the context of the development of space astronomy at TIFR,  I had sketched the exciting story of how TIFR used its know-how of balloon techniques to make inroads in space astronomy, indulge in many diverse activities, eventually culminating in the launch of India's first multi-wavelength astronomical satellite AstroSat in 2015, in which TIFR played a pivotal role 
 \citep[hereafter referred to as Paper1]{RaoSpaceTIFR}.

\subsection{Scope of the present paper}

Paper1 was based on a talk at a TIFR conference organized by the TIFR Alumni Association from December 17-18, 2022. Reflecting the `feel-good' mood of the meeting, Paper1 gave an extremely positive vibe to the events leading to the launch of AstroSat. 

The story of this exciting journey of a developing country to be a part of a front-line research area has several lessons to offer. 

Several streams of thoughts about doing science could be examined here. 
\begin{itemize}
\item
Development of an area of science: individual or cultural?
\item How did the corporatization of science, brought about by X-ray astronomy, influence the way of doing science?
\item Whether there is still a place for blue-sky research?
\item How can these insights be used to formulate a strategy to do X-ray astronomy in the future, particularly in India?
\end{itemize}

With these questions in mind, I have attempted to examine the development of X-ray astronomy in India. I have subjectively divided the past six decades of research work in X-ray astronomy into three broad eras. The era from the birth of X-ray astronomy in 1962 until 1980 saw a rapid expansion of X-ray astronomy. It coincided with a period of `dreamers' from India aggressively trying to catch up with this exciting journey of new discoveries. The next era, from 1980 to 2000, saw a consolidation of work in X-ray astronomy brought about by the Great Observatories. During this period, India was struggling to catch up. This millennium, however, brought about a maturity in X-ray astronomy, the primary driver being the thinking of niche observational areas of research that are topical and scientifically relevant. In India, this era saw a renaissance in X-ray astronomy, culminating in the launch of AstroSat. Finally, the future could be a balance between striving to leapfrog by vastly expanding observational sensitivities and developing specific science-driven observational campaigns. Here, I will dwell on some aspects of the style of doing science and will give some personalized suggestions about advancing X-ray astronomy as a collective creative art.

I must hasten to add that this article is not a rigorous and serious review of X-ray astronomy. My understanding of X-ray astronomy is based on the practice of this discipline for about four decades. I have gone through most of the articles on the subject in the Annual Review of Astronomy and Astrophysics for this article. I have referred to these articles at appropriate places but have refrained from giving an extensive bibliography. This has enabled me to retain an overall birds-eye view of the subject so that some general trends could be identified and used to obtain a reasonable future strategy, particularly in the Indian context.

Apart from the Annual Review articles, I benefited from the material available from the HEASARC website and also some good review articles and books 
\citep{PCAReview, SantangeloReview, GiacconiBook, Pulakkat}.

\section{1962 - 1980: rapid expansion; era of dreamers}

Once it is known that bright X-ray sources exist in the sky, the scientific community voraciously investigated the nature of these sources using all available resources to send equipment above the Earth's atmosphere. In the initial years, however, rockets and balloons were available as the carriers of the instruments. The short duration of exposure severely limited rocket-based experiments. The balloon-borne instruments, however, could reach heights only up to about 40 km above Earth, and the residual atmosphere at that height blocked most of the X-ray radiation below about 20 keV, resulting in relatively poor sensitivity for observations. Thus, by 1966, the total exposure to the sky was about an hour from rockets and $\sim$ 150 hours from balloons. About 20 X-ray sources were detected, and only about half were confirmed by multiple observations. A few sources, like Crab and Sco X-1, were securely identified. Most sources were found in the Galactic plane, mainly near the Galactic centre. There were claimed, but unconfirmed, detection of sources close to extragalactic objects like Coma cluster and M82. The ubiquitous X-ray background was identified to be of cosmic origin. Proportional counters in soft X-rays and scintillating crystal detectors in hard X-rays were the detectors of choice, and various ways like scanning collimators, modulation collimators and lunar occultation techniques were used to localise sources.   It was realised that grazing incidence mirrors are needed to focus X-rays, and a small area X-ray focussing telescope could only be used to image the Sun.    It was concluded that X-ray astronomy was a `strong if immature branch of science' and it could be `..one of the most fruitful channels of data about the largest scales of the universe', or it could be relegated  `to the detailed study of certain unusual stars'  \citep{Morrison1967}. In fact, at that time, very few precise measurements were available in X-ray astronomy, and it was realised that the field was in the `last stages of simple exploration' and could enter the era of accurate measurements through satellite-based observations \citep{Giacconi1968}.

 UHURU, launched in 1970, firmly established X-ray astronomy as a mature and modern tool to understand the cosmos. Though modest by current standards (just 142 kg of instruments), the design was very carefully done (Table 1). The detectors were two large area proportional counters with two different field-of-view collimators. The detectors were continuously rotated to scan the sky. The satellite was thoughtfully put in a near-equatorial orbit to achieve a low and stable background and avoid the high background South Atlantic Anomaly (SAA) region. The continuous observations possible in a satellite platform were a vast improvement compared to barely an hour of observation possible during the sum-total efforts of all previous rocket experiments in the low energy ranges  (1.7 - 18 keV). UHURU detected 161 sources in the first 125 days of operation, and 40 were identified. Detection of X-ray pulsations in two bright X-ray sources in the Galactic bulge (Cen X-3 and Her X-1) and measuring their binary periods firmly put accreting compact objects as the sources of bright Galactic sources. The precise X-ray position of Cyg X-1 led to radio identification, which in turn helped the optical identification and, later, in 1975, its binary nature \citep{Bolton1975}. The simplest explanation was an accreting black hole in Cyg X-1, though other possibilities could not be unequivocally eliminated \citep{BlumenthalTucker1974}. The bright Galactic sources like Sco X-1 and Cyg X-2, with no pulsations or clear evidence of binarity, remained mysterious.
UHURU  also detected  40 high latitude sources, 20 of them firmly identified with extragalactic objects like  AGN, QSOs, and clusters of galaxies,  thus making X-ray astronomy `..one of the most fruitful channels of data about the largest scales of the universe'.

\begin{table}[htb]
\caption{Important X-ray astronomy missions during 1962-1980$^a$}\label{tableExample} 
\begin{tabular}{lccl}
\hline
Name& Operation & Mass (kg)  &Instruments \\\hline
UHURU&1970-73& 142& Collimated Proportional Counters\\
& & & $~~~$(0.5$^\circ$$\times$5$^\circ$) \& (5$^\circ$$\times$5$^\circ$);  840 cm$^2$; 1.7 - 18 keV \\
HEAO-A&1977-79& 2552&A1: Scanning Sky Survey\\
& & & $~~~$ 2000 cm$^2$; 0.25 - 25 keV\\
& & &  A2: Collimated detectors (LED, MED, HED)\\
& & & $~~~$3000 cm$^2$; 0.15 - 60 keV \\
& & & A3: Scanning modulation\\
& & & $~~~$ 450 cm$^2$; 1 - 13 keV\\
& & & A4: Collimated detectors (LED, MED, HED)\\
& & & $~~~$ 220 cm$^2$; 15 keV  - 10 MeV \\
HEAO-B &1978-81& 3130 & High Resolution Imager (HRI)  \& grating \\
& & & $~~~$2"; 20 cm$^2$; 0.15 - 3 keV  \\
& & & Imaging Proportional Counter (IPC)\\
& & & $~~~$1'; 10 cm$^2$; 0.4 - 4 keV\\
& & & Solid State Spectrometer (SSS) \\
& & & $~~~~~~$ 200 cm$^2$; 0.5 - 4.5 keV\\
& & & Focal Plane Crystal Spectrometer (FPCS) \\
& & & $~~~~~~$ 1 cm$^2$; 0.4 - 2.6 keV\\
& & & Monitor Proportional Counter (MPC)\\
& & & $~~~$1.5$^\circ$; 667 cm$^2$; 1.5 - 20 keV\\
\hline
\end{tabular}
\tablenotes{$^a$Angular resolution, detector area and the energy range are given for each instrument}
\end{table}

The success of UHURU led to several small-scale experiments like Copernicus (OAO-3), ANS, Ariel-5, Ariel-6, and SAS-3. Accretion onto compact objects emerged as a mature field of study, and most of our current understandings of accretion had their seeds firmly planted in the seventies. The leap of faith to understand Active Galactic Nuclei as accreting supermassive black holes was yet to mature. While discussing extragalactic X-ray sources
\citet{GurskySchwartz1977} concluded that the unanswered questions are  
``of origin and evolution which we are yet too timid to ask''. \citet{Weedman1977}, while discussing Seyfert Galaxies, opined that they are similar to QSOs, but ``...the reason for their existence remains one of the most pressing astrophysical mysteries''. 
Inspired by stellar black hole sources, they also discussed the possibilities of having   10$^7$ M$_\odot$ black holes or neutron stars  but concluded that 
``..how the required compact objects get into the nuclei requires comprehensive knowledge of
 galactic evolution...'' 
 
 X-ray astronomy came into full maturity after the launch of HEAO-A in 1977 and HEAO-B (Einstein Observatory) in 1978 (see  Table 1). With about 3000 kgs of instruments in each of these two satellites, they are on par with modern large-scale experiments in scale and vastness. HEAO-A pushed to the limit the capabilities of non-imaging X-ray telescopes with a collecting area close to a square meter. The advantages of using focussing techniques in X-ray telescopes are humongous. The ubiquitous X-ray background can be tamed. Firstly, the `seeing' for a source is in arc seconds compared to the degree scale achievable in collimated detectors, thus reducing the solid angle for the isotropic background by a million times. Secondly, the geometric area of the detectors is at least a couple of orders of magnitude lower than the collecting area, thus reducing the internally generated background. Lastly, the imaging properties of the detectors can be gainfully employed to measure and correct for the residual background simultaneously.
Further, the small detector area required to measure X-rays from a large collecting area offered by the X-ray mirrors afforded the luxury of developing innovative high-tech X-ray detectors at the focal plane. Einstein Observatory used a whole gamut of possible new X-ray detector concepts. These included arc-second imaging capabilities, good resolution solid state detectors and high-resolution spectrometers (see Table 1).

The results from these two observatories fully justified the expectations. X-rays from clusters of galaxies were fully understood \citep{FormanJones1982}, and X-ray emission was detected from all classes of astronomical objects ranging from stellar coronae to distant quasars. Complementary X-ray observations have become a necessity to understand any high-energy astrophysical phenomena.

\subsection{Meanwhile, in India...}

The fascinating journey of X-ray explorations excited and enthused scientists across the globe. The newly established scientific institutes in India, like the Tata Institute of Fundamental Research (TIFR) and Physical Research Laboratory (PRL), made it a point to expose young researchers to new scientific activities by sending them to Western countries, particularly the USA. These young researchers, whom I call the dreamers, with their unbridled enthusiasm, were responsible for India's foray into X-ray astronomy.

B. V. Sreekantan was one of the first students of Homi Bhabha, the founder of TIFR. Sreekantan was a trained experimental Cosmic Ray physicist. He was at MIT during the sixties and was fortunate to witness the fascinating story of the birth and growth of X-ray astronomy. At that time, Homi Bhabha had collected many young and enthusiastic scientists, and the mood and ambience at TIFR were one of attacking any scientific problems with a sense of adventure. TIFR had a vigorous experimental activity of studying cosmic rays using balloon-borne instruments. Sreekantan used this infrastructure to start a balloon-borne X-ray program at TIFR.  
 He also played an essential role in motivating and training many students. In the initial years, he proactively participated in balloon-borne X-ray astronomy experiments to get exciting results like the discovery of `Sudden Changes in the Intensity of High Energy X-Rays from Sco X-1'  \citep{Agrawal1969Sco} and the detection of `Rapid Variations in the High Energy X-ray Flux from Cyg X-1'  \citep{Agrawal1971Cyg}. After he became the Director of TIFR in 1975, he was the perfect benevolent fatherly figure to nurture and support all the new initiatives. He also navigated and managed the socio-political ambience to support these new activities adequately.
 
 Similarly, U. R. Rao, one of the first students of Vikram Sarabhai, the founder of PRL, was in the USA in the sixties studying Cosmic Rays. He, too, was suitably impressed by the new developments in X-ray astronomy and made it a point to start balloon-based and rocket-based X-ray astronomy experiments at PRL. He was primarily responsible for the rocket-based X-ray astronomy experiments in India, and several exciting results on the spectral and timing studies of X-ray binaries like Sco X-1 and Cen X-2 were obtained  \citep{Rao1969a, Rao1969b}. PRL later morphed into an Institute under the umbrella of the Indian Space Research Organisation (ISRO), and U R Rao moved to ISRO and played a pivotal role in the early initiatives of ISRO in X-ray astronomy. 
 
 India enthusiastically developed several X-ray instruments during the post-UHURU era when small-scale satellite experiments were attempted. 
 Aryabhata, the first indigenous Indian satellite launched in 1975,  had an X-ray astronomy experiment (though the satellite worked only for about a few days). Bhaskara satellite, launched in 1979, was mainly a remote sensing and communication satellite but had an  X-ray sky monitor (lasting only a few days of operation).    Though the initial instruments were based on crystal scintillating detectors for studying cosmic X-ray sources, new space X-ray experiments based on proportional counters were initiated later. These included a rocket-based investigation to explore the soft diffuse X-ray background and a large Xenon-filled proportional counter for a balloon-borne hard X-ray experiment.

\subsection{A critical summary of the first era of X-ray astronomy}

I will first outline the theme of the two ways of doing science: `corporate' style and `classical' science.

In contrast to `corporate science', where method, efficiency and end results are essential and quantified,  `classical science' is the sort of blue-sky research driven by individual brilliance where the process of doing science is essential. The results in the former are estimated in advance, and the efficiency quantified, whereas in `classical' science, the results are only a by-product of this way of doing science. Product is essential for the former, whereas process is vital for the latter.

The corporatisation of science had its seeds in the organised efforts to do particle physics using accelerators at CERN, founded in 1954. The growth of X-ray astronomy clearly demonstrates corporate governance's utility in large-scale science activities. From the rudimentary detectors used in the discovery rocket flight in 1962, growing as a mature field in less than two decades is truly remarkable. Of course, the vast amount of resources available in that era was also a tremendous positive incentive. Still, the emergence of a new style of doing research was required to bring about this growth.

 What are the negative aspects of corporate science?
 
  For one thing, only problems and targets amenable to a well-defined articulation would get precedence. One needs more prolonged exposure to make better measurements in X-ray astronomy: good, build a satellite. One needs grazing incidence X-ray focussing mirrors to improve sensitivity: excellent, build one. In this style, subtle questions for a deeper understanding of science, which need open-ended explorations, will lose out and, in a long-term perspective, could be detrimental to the growth of science.
  
   Are there any areas in X-ray astronomy which did not get sufficient support due to the corporatisation of science?
   
    Yes, I believe so. Let me articulate the concepts which will be developed later on.

Astronomy is essentially measuring the Astros and making sense of them. The vastness of the universe, in space and time, resulted in islands of matter settled in equilibrium for long durations of time, like the stellar interiors. The knowledge of physics and precise measurements in the optical band allowed a detailed understanding of such steady-state objects in equilibrium - I will call this `thermal' astrophysics. Of course, the observation wavelength might focus more on a specific temperature band (like the optical band being sensitive to the temperature range of stellar surfaces). Observations in many wavebands more or less covered all temperature ranges that could be conceivably present in the universe in steady states, ranging from the microwave background to the surfaces of hot white dwarfs. 
     
    In contrast to such steady state `thermal astrophysics', observations during the past several decades have shown that there exist certain `non-thermal' astrophysical phenomena where energy is transferred from one form to another in a non-equilibrium fashion, ranging from the magnetic reconnection events in solar flares to the humongous release of the rest mass energy of one solar mass in a few seconds during the birth of a compact object (or coalescing of two compact objects) resulting in Gamma-ray Bursts (GRBs). Additionally, most non-thermal phenomena we observe today are closely coupled to physical phenomena near black holes (in GRBs, AGN, and X-ray Binaries). Hence, a detailed understanding of such phenomena is expected to expand our knowledge of matter in extreme conditions (high magnetic field, high gravity, etc.) and possibly lead to new physical principles and understandings.
     
       The manifestations of non-thermal phenomena have observables in different wavelengths. The energy is mainly transferred to electrons; the immediate manifestation is non-thermal hard X-ray emission. The protons and other nuclei can be accelerated to high energies, giving gamma-ray emissions. However, the cooling time scales are much longer than electrons; hence, they are a delayed manifestation of the putative energy trigger. When the energy loss rate is much lower, these accelerated electrons can remain energetic for a long time. Any ambient magnetic field can make these electrons emit in radio wavelengths - allowing the study of the remnants of non-thermal phenomena.
       
       The   X-ray range is in the cusp of the thermal and non-thermal emissions. A detailed understanding of the non-thermal emission and the coupling of the thermal and non-thermal emission is hampered by the lack of improvements in the hard X-ray sensitivity. Low energy X-rays, up to 6.4 keV (the Fe K-alpha line), are primarily thermal emission and improvement in the sensitivity of X-ray telescopes, though difficult, is straightforward: build better and larger mirrors. Hence, over the past six decades, we have seen almost a million times improvements in the sensitivity of X-ray telescopes in soft X-rays. The improvement in hard X-rays, particularly above 50 keV, is marginal.
       
        The importance of hard X-ray measurements has been known for a long time, but the avenues to improve them are not very clear, nor are they straightforward. The point I am making here is that the `corporate culture' of doing science is not amenable to attacking such open-ended problems. 
        
          I will expand on these ideas and try to quantify them later in this article.

\subsubsection{The Indian context}

 If one wants to summarise the efforts in India during the first era of X-ray astronomy, India benefited from its emphasis on the `classical style' of doing science to get into X-ray astronomy. However, it failed to make an impact essentially due to her inability to get into the corporate style.
 
  The foray into balloon-borne and rocket-borne X-ray astronomy in India was done when it was realised that the field was in the `last stages of simple exploration' using balloons and rockets \citep{Giacconi1968}. Though it is commendable that significant results on the spectral and timing properties of X-ray binaries were obtained, 
  these results were of the type to find a reference in a review of the individual sources but are of little value to find mention in a review of a broad research area of X-ray astronomy. 
  Additionally, in the area of instrumentation in X-ray astronomy, there was not a single instance of significant discoveries or improvements made in India that merited an original contribution. 
  I will examine some sociological aspects that could have prevented India from substantially contributing to X-ray astronomy in this era. 
  
  In this era, the primary centre for X-ray astronomy was TIFR (with some contributions from PRL and, later, in satellite-borne X-ray astronomy, ISRO). Hence, I will primarily concentrate on working with the X-ray astronomy group at TIFR. 
  
  The style of working of many scientific Institutes in India in this era, like TIFR and PRL, was driven by a sense of doing deep science; the icons of science are the great scientists of the early twentieth century, exemplified by Albert Einstein. The founding logic of these Institutes is quite simple: science in the nineteenth and early twentieth centuries was driven by brilliant individual scientists. Collect the best and brightest from all across India, nurture and encourage them; surely, the result will be great science.

 There is an underlying assumption in the basic premise of the above method. The assertions that science was driven by `brilliant individual scientists' and if you collect the `best and the brightest' you can nurture them to be `great scientists' presupposes that there exists a separate class in human talent, distinctly different from the average, destined to be great, only the circumstances prohibiting them becoming great like the superlatively talented individuals born during the birth of India ending up as wastrels in the poor Indian conditions, as imagined in the novel `Midnights Children' by Salman Rushdie. Quite often, the marks of students appearing in the entrance test of TIFR were carefully examined to look for two Gaussians in the distribution: the broad lower one representing the vast dumb majority and a sharp short peak on the right side representing the talented individuals. 
  
    However, alas, the distribution was mainly a power law.

Hence, I want to advance the following hypothesis to make sense of the development of X-ray astronomy in India.
     
  Instead, the inherent human talent in a population could be a Gaussian, like many other measurable quantities like height, weight of the brain, etc. The distribution can become a power law when talent growth requires multiple social interactions. 
     In a settled society, encouraging the high end of the power-law distribution in talent may help develop a more mature community. In a poor country like India, the education system was primitive in the fifties and sixties. Typically, in a given class in an elementary school, only one or two students, fortunate to have positive family support, managed to pass the exam, and the rest dropped out of the education system. Hence, the obsessive search for extreme talents naturally ended up selecting sort of `freaks': people who concentrated on their studies despite great odds, people who had the willpower for extreme self-denial so that their studies would be strengthened at the expense of other social interactions. TIFR of the sixties and seventies was full of talented individuals, but most were socially awkward, self-centred and had obsessive egos. I felt that the general ambience was quite `caustic'. When I confided this to one of my senior colleagues, he said that the ambience was definitely `not healthy for broad based collaboraration' (S N Tandon, private communications).
    The final effect of a collection of talented but socially awkward individuals at TIFR was that there indeed was individual brilliance but very little collective activity. 

 The development of X-ray astronomy in India, particularly in TIFR, can be explained based on the above perspective. 
 Selecting talented individuals from across India and giving them training, facilities, exposure, and total freedom resulted in many accomplished scientists. 
  In a few years, India could produce very competitive results in X-ray astronomy and later enthusiastically get into satellite-based X-ray observations in the post-UHURU era.  
  Nevertheless, one cannot help noting that in TIFR of that era, the X-ray astronomy group had the ``best and the brightest from all across India'' with good support, but, sadly, the result was not great science.

    I strongly feel that the negative social aspects of a bunch of talented individuals prevented the nucleation of a strong group of people and, hence, prevented the development of a sort of  collective collaborative culture to attack deeper problems at TIFR.  
  Hence, new methods to make a mark in hard X-ray astronomy could not be evolved. In the post-UHURU era, there was a fertile scientific area of hard X-ray astronomy to be explored from the homegrown balloon facility at TIFR. Indeed, once there were three parallel hard X-ray balloon-borne experiments simultaneously carried out at TIFR in this era. Hence, though there were foreign collaborations for hard X-ray astronomy (with Japan, the USSR, and Canada), a strong group collaborating on an equal footing was lacking. Therefore, a substantial homegrown area of expertise was not developed.

\section{1980 - 2000: consolidation; the era of strugglers}

 The post-Einstein Observatory era of X-ray astronomy saw the consolidation of the gains made earlier. 
There were sustained efforts from  Europe, the USSR,  Japan, and the USA. The developments were quite straightforward: make collimated detectors based on proportional counters and scintillating crystals and improve the focussing techniques at lower energies. Since the X-ray sky is highly variable, all-sky monitoring was also one of these satellites' key ingredients.

\subsection{Consolidation}

From the European side, EXOSAT (launched in 1983; satellite weight 510 kg -  these two numbers would be given hereafter while introducing an X-ray mission) focussed on making wide-band spectral observations. It used the concept of having highly elliptical long satellite orbits to get uninterrupted long observations (91-hour period). Focussing X-ray mirrors at low energies  (a few cm$^2$ area) and a reasonably large area (1800 cm$^2$) proportional counters (along with the new concept of gas scintillation proportional counters to improve the energy resolution)  were the hallmarks of the detectors. ROSAT (1990; 2400 kg) had a much larger area of X-ray mirrors (400 cm$^2$), and with its large field of view  (2 degrees), it allowed a sensitive all-sky survey at low energies. The  BepoSAX satellite (1996; 1400 kg) had a modest X-ray mirror at low energies (50 cm$^2$ area), the widest band spectroscopic detectors and a Wide Field Camera.

The USSR's developments initially centred around space station modules, like the  Kvant module and the  Roentgen X-ray Observatory. Though a gamma-ray observatory, the Granat satellite (1989, 4400 kg) had proportional counter-based X-ray detectors (2400 cm$^2$) and a hard X-ray all-sky monitor.

Japan made a systematic effort to make bigger and better telescopes in the eighties and the nineties. Starting with Hakucho (1979, 96 kg), which is a modest UHURU class satellite with proportional counters  ($<$ 100 cm$^2$ area) and modulation collimators,  X-ray astronomy missions with increasing complexity and sophistication were attempted every few years. Tenma (1983; 216 kg)  had a gas scintillation counter and a modest area X-ray mirror (15 cm$^2$ area).   Ginga (1987, 400 kg) used very large area proportional counters (4500 cm$^2$) and an all-sky monitor (70 cm$^2$) to make timing measurements of X-ray sources. ASCA (1993, 420 kg) used the replicated X-ray mirror foil technique to get a large area  (1200 cm$^2$) at the expense of seeing, and it was the first X-ray astronomy mission to use X-ray  CCDs.

The USA used the Space Station to demonstrate replicated X-ray mirrors to get large areas (several hundred cm$^2$) in a very wide bandwidth. It used Si(Li) detectors to make good energy resolution spectral observations in the BBXRT (1990) instrument. RXTE (1995, 3200 kg) and the  Chandra Observatory (1999, 5000 kg), one of the Great Observatories of NASA, were significant contributions from the USA in this era.

 The superior sensitivity and longevity of Chandra, coupled with a similar set of properties of XMM-Newton along with the socially impactful RXTE (see Table 2), made X-ray astronomy a complete science.

\begin{table}[htb]
\caption{Major  X-ray astronomy missions during late ninties}\label{Mission90} 
\begin{tabular}{lccl}
\hline
Name& Operation & Mass (kg)  &Instruments \\\hline
RXTE&1995-2012& 3200& Proportional Counters Array - PCA (6500 cm$^2$)\\
& & & $~~~$ 2 - 60 keV\\
& & & All Sky Monitor - ASM  \\
& & & HEXTE (1600 cm$^2$) \\
& & & $~~~$ 15 - 250 keV\\
Chandra&1999- & 5000&High Resolution Camera (HRC)\\
& & & $~~~$$<$0.''5;  200 cm$^2$ at 1.0 keV \\
& & &  Advanced CCD Imaging Spectrometer (ACIS)\\
& & & $~~~$ 1";  600 cm$^2$ at 1.5  keV\\
& & & Transmission Grating Spectrometers\\
XMM-Newton &1999- & 3700 &European Photon Imaging Cameras (EPIC) \\
& & & $~~~$1"; 4425 cm$^2$ at 1.5 keV   \\
& & & Reflection Grating Spectrometers (RGS)\\
& & & Optical Monitor (OM) \\
& & & $~~~~~~$ 30  cm dia telescope \\
\hline
\end{tabular}
\end{table}

\subsection{The science context}

 As a mature branch of astronomy, X-ray astronomy contributed to all areas of astrophysical observations. 
At this stage, let me subjectively divide the research areas in astrophysics,  primarily driven by X-ray observations.

\begin{itemize}
\item
{\bf X-ray Binaries:} Bright Galactic X-ray binaries (XRB) were, traditionally, the prime focus of study in X-ray astronomy. These accretion-powered compact objects (neutron stars or black holes) primarily emit X-rays. A detailed study of the spectral and the timing properties of these objects is expected to yield answers to very fundamental astrophysical questions like a) measuring the fundamental parameters of compact objects (mass and spin of black holes, and, additionally, size and magnetic field of neutron stars and b) study of  Physics in extreme conditions near compact objects.
\item {\bf X-ray Spectroscopy:}  There are many classes of objects, which are copious X-ray emitters, including stellar flares and coronae, cataclysmic variables, Active Galactic Nuclei, clusters of galaxies, etc. A detailed spectral study of many such bright (a few mCrab fluxes) X-ray sources was possible due to X-ray focussing mirrors.
\item {\bf Deep X-ray imaging:} Deep observations by high-quality X-ray imagers like Chandra can go very deep and provide complementary observations for any astrophysical object. 
\end{itemize}

\subsubsection{X-ray Binaries  }

Masses of compact objects were estimated in about six X-ray binaries before the HEAO era \citep{Bahcall1978}.

 Though neutron stars are known as the sources of radio pulsars, `relatively little can be learned about neutron star masses or radii' from the radio observations 
\citep{JossRappaport1984}. X-ray observations in the HEAO era increased this number significantly. By 1990, there were 100 known binary neutron star systems (81 of them being X-ray binaries), and several properties of neutron stars were well documented \citep{Canal1990}.
One of the very significant contributions of the EXOSAT satellite is the detection of Quasi-Periodic Oscillations (QPO) and red noise in neutron star XRBs 
\citep{vanderKlis1989}. 
Ginga X-ray spectroscopic measurements detected the cyclotron resonance features in several sources 
thus measuring their magnetic fields.
Tenma satellite identified the two component spectra in LMXBs and absorption lines at 4.1 keV in X-ray bursts \citep{TanakaShibazaki1996}.

Evidence for BH in XRBs was initially confirmed for only three sources (Cyg X-1, LMC X-3, and A0620-00). The number significantly increased due to the presence of all-sky monitors in the eighties and nineties.
By 1996, there were 124 known LMXBs; 41 of these were transients, including 18 BH LMXBs (all BH LMXBs are transients), and 
 radio outbursts were seen in 8 of them,  though there was no talk about jet emission in X-ray binaries till the early nineties
 \citep{TanakaShibazaki1996}.
QPOs were noticed in BH candidates in the EXOSAT era, but `the full pattern, if there is one, has not yet emerged'   \citep{vanderKlis1989}. 

The launch of the RXTE satellite in 1995 made a dramatic and impressive impact on the study of XRBs. Though the prime detector, the Proportional Counter Array (PCA), 
is similar to the proportional counter-based Large Area Counter on the Ginga satellite launched in 1987 (about 20\%  better sensitivity), RXTE was in the internet era. RXTE which included 
PCA, the All Sky Monitor (ASM), and the hard X-ray detector HEXTE had a wide energy range, excellent timing accuracy (1 $\mu$s), and the flexibility to target any desired object quickly. 

Nevertheless, the most impressive aspect of RXTE is the vibrant use of the internet to co-opt a large section of the astrophysical community to go deep into the problems thrown open by the new observations. Though the USA had a vibrant `guest observations' and `archival data analysis' program starting from the HEAO satellites, the establishment of 
 HEASARC  in  1990 at NASA's Goddard Space Flight Center (GSFC) with an explicit charter to support high-energy astronomy in all aspects of data analysis and the wide use of the internet profoundly affected the use of the data from RXTE. Additionally,  observations of bright XRBs, some newly discovered transients, generated vast spectral and timing data on these highly variable sources. The easy availability of such highly variable  (and potentially scientifically new) data enabled the emergence of a considerable number of `entry' level scientists (`plot a light curve and write a paper'), thus generating a vast but perhaps quite elementary, bottom level in the pyramid of knowledge, based on whom, quite conceivably, higher levels of scientific endeavours could be built.   I will say RXTE formalised the concept of writing scientific papers as an educational tool.
 
  Scientifically, RXTE proved to be highly rewarding. New phenomena at  millisecond time scales were discovered  in X-ray binaries:  millisecond pulsations, burst oscillations, and kHz QPOs
\citep{vanderKlis2000}. 
Though relativistic jets were discovered in the Galactic X-ray binary GRS 1915+105 (discovered by the Granat satellite) before the launch of RXTE, the all-sky monitoring ability, quick observation sampling and easy availability of data from RXTE enabled the discovery of many relativistic jet sources and their detailed understanding 
\citep{MirabelRodriguez1999}. 
GRS 1915+105 proved to be an extremely fascinating object to study. It had several episodes of superluminal jet emissions and multiple types of variability classes, and it was a transient source that `forgot to fade away'. This source merited its own separate article in the Annual Reviews \citep{FenderBelloni2004}. 
There were 20 BH-XRBs with mass measurements, all monitored by RXTE, and almost 
 half discovered by RXTE \citep{RemillardMcClintock2006}.
Such extensive studies helped connect the accretion phenomena in XRBs and AGNs by discovering fundamental relationships between X-ray and radio intensities and masses of black holes across black hole mass ranging from stellar mass to supermassive. The quick pointing ability of RXTE helped very long-term monitoring of bright AGNs, enabling the discovery of the dependence of the 
break time scale in the power spectral density to the mass and luminosity across mass scales   \citep{McHardy2006}. 
It was established that the `physics of disc-jet coupling in XRBs and AGN are closely linked' \citep{FenderBelloni2004}. 

\subsubsection{X-ray spectroscopy }

Though X-ray spectroscopy using EXOSAT, Ginga and Tenma established several characteristics of `faint' (a few milliCrab) sources like the 
absorbed X-ray spectra of Seyfert 2s, universal power-law in AGNs, along with a hint of reprocessed reflection component above eight keV  (Ginga), it was the superior medium resolution 
X-ray spectroscopy using the XMM-Newton EPICpn instrument and the high-resolution grating spectra (Chandra and XMM-Newton complementing each other to get wider bandwidth) that 
X-ray spectroscopy finally fully matured \citep{Frederik2003}.
In the coronae of late-type stars, the FIP effect on abundance was inferred, and the Doppler effect on the spectral lines indicated that emission is mainly near the poles.
High-temperature gas was detected in early-type stars. Although spatially resolved high-resolution spectroscopy could not be done in SNR, high-resolution spectra coupled with the spatially resolved medium-resolution spectra could provide insights into these objects' temperature and abundance distribution. The excellent quality spectra on hot White Dwarves enabled very detailed modelling.
P Cygni profiles were detected in XRBs and CVs. 
A detailed study of clusters enabled the understanding of the intricate dynamical evolution of these sources.

\begin{figure}[h]
\centering
\includegraphics[angle=-90, width=\textwidth, trim={2cm 1cm 6cm 3cm},clip]{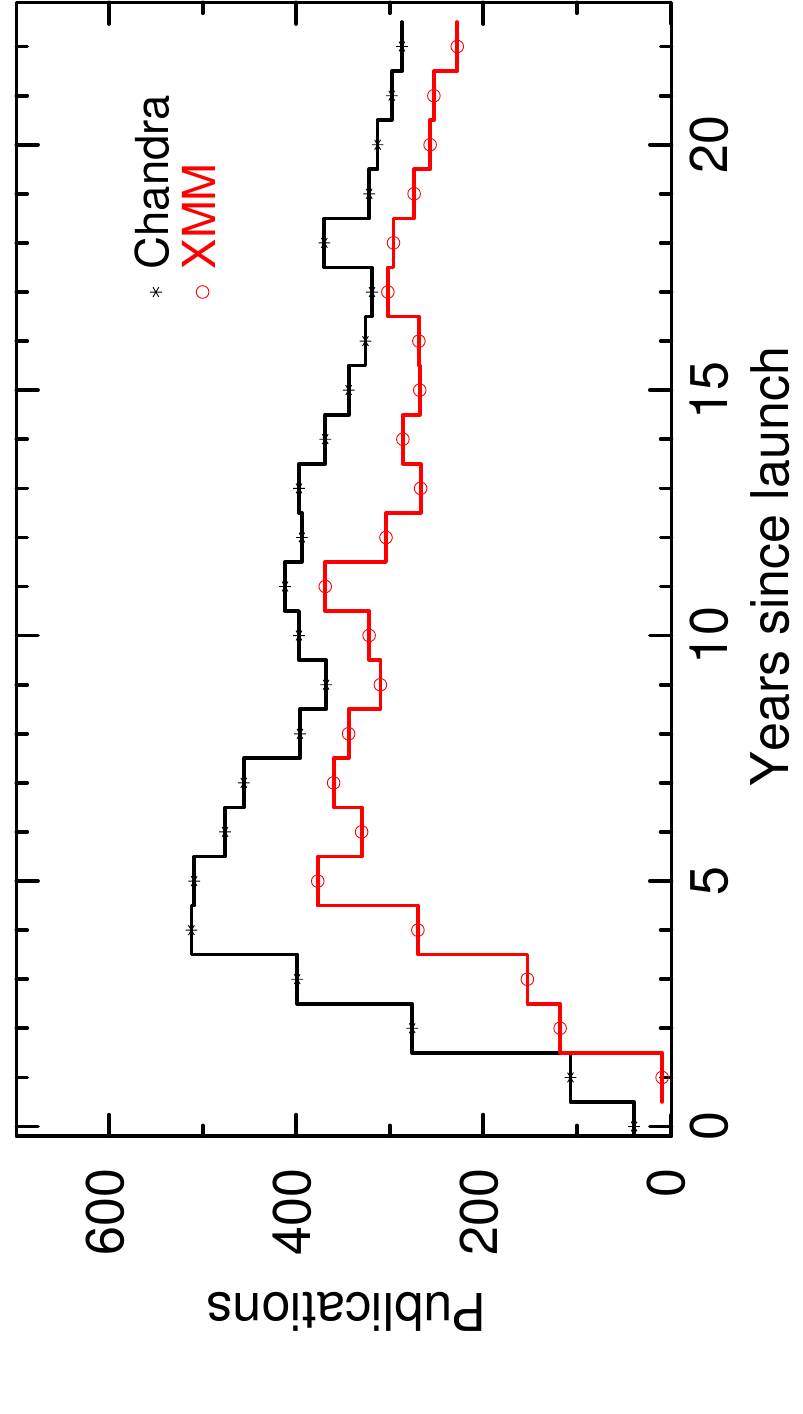} 
\caption{Number of refereed publications based on data from  Chandra Observatory and XMM-Newton (both launched in 1999).}
\label{fig1}
\end{figure}

\subsubsection{Deep Imaging}

The deep imaging provided by Chandra brought about a revolution in sensitivity \citep{McCammonSanders1990}. Though the study of clusters of galaxies began with the detection of a few clusters with UHURU and the detection of a  few dozen during the HEAO era, deep imaging by Chandra and medium-resolution imaging spectroscopy by XMM-Newton enabled doing cosmology with X-ray clusters by detecting them up to a redshift of 1 and beyond and discovering hierarchal model of structure formation
 \citep{Rosati2002}.
Deep extragalactic X-ray surveys using Chandra and XMM enabled going deep by one to two orders of magnitude deeper than ROSAT and allowed 
the X-ray measurements of AGN evolution and the growth of supermassive black holes, providing constraints on the Physics of high-redshift (z $>$ 4) AGN 
  \citep{BrandtHasinger2005}.
The fluctuation analysis in the X-ray images enabled going deep by another couple of magnitudes fainter. 
 Deep imaging also allowed the detection of X-ray binaries up to the Virgo Cluster, the study of the population of X-ray sources in galaxies, including the Ultra-luminous X-ray sources \citep{Fabbiano2006}, and the detection of 
X-ray emission from extragalactic jets \citep{HarrisKrawczynski2006}.

\subsection{In India, struggling to catch up ...}

 X-ray astronomy was initially done at very few places in India: TIFR and PRL were the only two places. After 1980, many X-ray astronomers from PRL moved to ISRO and took up higher responsibilities like building Institutes for satellite and rocket fabrication. Though some work on Gamma Ray Burst (GRB) observations and optical observations of X-ray sources were carried out at PRL and ISRO, mainstream X-ray astronomy in India in the eighties and the early nineties was primarily done at TIFR. Hence, I will concentrate on the sociological aspects of doing X-ray astronomy at TIFR.

Let me repeat: TIFR started with the philosophy of collecting the best and the brightest and giving them ample resources and support to deliver great science.

But, sadly, the result from TIFR was not great science.

Earlier in this article, I articulated the possibility that collecting `best and the brightest' and supporting them may not be sufficient conditions for great science. I  advanced the possibility that other ingredients, like the ability to work collectively, may be required.  

Since the founding philosophy of TIFR was attributed to Homi Bhabha, the great founding director of TIFR, in the early eighties, the effort was to scrutinise the lack of good quality science under the sanctity of the founding principle. Hence, the diagnosis and the remedy veered on elementary premises. Like, since the outcome is not good science, most likely, the collected individuals are not bright enough for the task.

 There were efforts to weed out the perceived laggards. There were attempts to make the entry qualifications stricter. The practice of selecting the available best and giving them jobs was abandoned: it was perceived to encourage inbreeding. 
 
 The combined effect of such remedies on the working of scientific groups at TIFR, particularly those requiring extensive collective efforts like X-ray astronomy, was disastrous. 

The number of X-ray astronomers working in TIFR was about ten when I joined TIFR in the late seventies (including freshly appointed persons like me and veterans like B V Sreekantan), and this number was reduced by half within a decade. Additionally, there was minimal incentive for collective activity. Everyone had to prove their mettle as science leaders. Hence, eventually,  the few individuals at TIFR ended up doing research in a few diverse areas with little cohesion and coordination. There were a few independent balloon-borne experiments for hard X-ray astronomy. There was also a collaboration with the Soviet Union called the Gamma-ray Indo-Soviet Program (GRISP). 
Archival or proposal-based X-ray data analysis was done using data from several observatories like HEAO-A, Einstein Observatory and EXOSAT.  

Finally, when the Polar Satellite Launch Vehicle (PSLV) capable of sending 2000 kg satellites to near-Earth orbits was developed and matured, X-ray astronomers from TIFR collaborated with ISRO  to make  the 
Indian X-ray astronomy experiment, IXAE.

\subsubsection{The Indian X-ray astronomy experiment, IXAE}

The Indian X-ray astronomy experiment (IXAE) was a small UHURU class satellite which included a proportional counter-based instrument called Pointed Proportional Counters (PPC) and one pinhole camera-based X-ray Sky Monitor  (XSM). IXAE was developed in India and launched (from India) in 1996 \citep{PCAReview}.  

The story of how TIFR and ISRO collaborated to make this instrument quickly and get exciting results on bright XRBs was told in Paper1. Here, however, I will concentrate more on the sociological aspects of this effort. Several exciting, positive aspects of doing X-ray astronomy fortuitously came together for this effort, and I will try to expand on these.

\begin{itemize}
\item
{\bf Experienced talent in X-ray astronomy:}
Though I was lamenting earlier the dwindling support for large-scale experiments like X-ray astronomy at TIFR, the ethos of TIFR, namely, giving total freedom to individual scientists and having an inherent expectation of getting `great' science from them, though had the effect of curtailing group efforts, encouraged individual scientists to pursue their passion. There were efforts to use modern data analysis techniques to understand topical astrophysical problems. There were efforts to understand the results of homegrown experiments by coupling them to results from other observatories and pitting them against the available current models. There were efforts to take up new techniques in X-ray instrumentation. There were continued efforts to persuade science management, particularly from ISRO, to invest in large-scale X-ray astronomy experiments. Though there were no `great' scientific results, there was indeed a sufficient amount of `good' science, and the scientists were experienced enough to take up the challenge of making an X-ray astronomy instrument.
\item 
{\bf A pool of technical expertise:}
Science, in general, and X-ray astronomy, being no exception, is a creative art. Good talent and continuous growth of this talent are essential for doing good science. On the other hand, for experimental science, particularly those needing large-scale experiments like X-ray astronomy, making instruments work reliably and well requires a different set of qualities, like perseverance, attention to detail for the subtle outcomes from the experiments,  following specific tasks, which could be monotonous and repetitive, with a single-minded devotion to bring them to a conclusion etc. TIFR did have a pool of technical people who had a lifetime of experience making instruments work for X-ray astronomy. Though no patents were in their name, they could make an amplifier work, for example.
\item 
{\bf Islands of technical excellence:}
TIFR concentrated on encouraging talent to pursue  good science. ISRO, on the other hand, had to get things delivered. Particularly after the early setbacks for rocket development in the eighties, there was a solid determination to concentrate on quality and make things work. One of the very good offshoots of this is the emergence of pools of technical excellence at ISRO in various space technology areas. A strong collaboration between TIFR and ISRO for IXAE helped tap these resources and make the instruments work.
\item 
{\bf Bits and bytes:} 
The establishment of HEASARC in the early nineties, the launch of RXTE in 1995, and the consequent easy availability of data and software galvanised the use of IXAE data. TIFR scientists were already well versed in using international data and implementing specialised software to homegrown instruments. This enabled an easy adaptation of analysis techniques to IXAE data.
\item
{\bf Science and its growth:}
HEASRAC and RXTE made X-ray astronomy data more easily accessible and made doing science in X-ray astronomy more democratic. This helped X-ray astronomy grow in India beyond the traditional strongholds like TIFR (and PRL/ ISRO). Many students across Indian universities and scientific Institutes became interested in X-ray astronomy and started interacting with TIFR.
\end{itemize}

The happy congruence of all these made IXAE an extremely positive impact on the growth of X-ray astronomy in India. Though the data from IXAE were of limited quality compared to that from RXTE, the combined effect of using home-grown data in conjunction with the  RXTE data made the science contribution very meaningful. Further, RXTE had opened up the field of the study of bright XRBs. While writing this article, I went through most of the related articles in the Annual Reviews and kept an eye on the number of times Indian work on X-ray astronomy was referred to by these articles. They were few and far between (at best, one or two results in some articles). Compared to this, nine IXAE and related results from India were mentioned in an  Annual Review article (Fender \& Belloni  2004). 

IXAE also had the effect of boosting the morale of Indian scientists: complex space instruments could be made, results could be competitive, and, more importantly, in my perspective, a large number of scientists could work together.

\subsection{A critical summary of the second era of X-ray astronomy}

\subsubsection{The science context}

Space astronomy, in general, is very expensive. The primary funding source is from the resources poured into space research, primarily driven by a sense of national pride. It is a folklore that the perceived loss of face by the USA  due to the dramatic and stupendous foray into space research by the USSR in the late fifties, sending an astronaut into space within a few years of the first artificial satellite,  resulted in the high resource-intensive Apollo program. X-ray astronomy benefited hugely from this and advanced as a mature branch of astronomy in the first era of X-ray astronomy, as detailed earlier. During the second era, however, the power of the USSR was waning, and the USA was recovering from the Challenger disaster, thus reducing the interest of these superpowers to pour higher resources into space research. Though the USA continued its commitment to developing the great observatory, Chandra (launched in 1999), the primary resource allotment to X-ray astronomy came from Japan and Europe, their national pride urging them to go ahead and be on par with the achievements of the USA in the previous era.

 Thus, the developments are incremental and based on specific improvements to the HEAO legacies, like EXOSAT providing continuous observation, ROSAT giving a good low-energy sky survey, and ASCA providing good spectral resolution. Though these observations consolidated the knowledge provided by the HEAO missions, there were no startling developments. Though it was known that hard X-ray astronomy was important, apart from the incremental efforts to make wide-band spectroscopic detectors like those in BeppoSAX, the number of known hard X-ray sources above 80 keV did not go beyond what was learned from the HEAO-A4 survey. As mentioned by 
 \citet{HeckmanBest2014}, while discussing the coevolution of galaxies and supermassive black holes, ``X-ray data having the depth and wide-field sky coverage to fully exploit the SDSS galaxy sample do not exist''. Hence, before the launch of Chandra and XMM-Newton, X-ray astronomy, by and large, was relegated `to the detailed study of certain unusual objects' like XRBs and AGN, rather than being  `..one of the most fruitful channels of data about the largest scales of the universe'.
 
\subsubsection{And in India...}

Research in X-ray astronomy in this era is primarily based on long observations from satellite-borne instruments. India did not have any presence in this area during this era. Of course, as I repeatedly hinted, science could have benefited by concentrating on niche observing areas like new developments in hard X-ray astronomy. However, the Indian science community was still in an individualistic mode of work to make any effort in this direction.

\section{2000 onwards: maturity; the era of renaissance }
 
 The launch of Chandra and XMM-Newton in 1999 and their continued operation till today, along with RXTE, which lasted for 17 years, fundamentally impacted X-ray astronomy. Even after two decades, Chandra and XMM-Newton are sustaining an impact on X-ray astronomy (see the number of publications using data from these two satellites in Fig 1). This was also the era of 
 INTEGRAL (2002, 4000 kg), a gamma-ray observatory having significant X-ray sensitivity, operating till now, and 
 Suzaku (2005,  1700 kg), a suit of the most sensitive wide band X-ray spectroscopic detectors, operating for a decade. With these advanced and sophisticated instruments, X-ray astronomy became a mature area of research. The impact of RXTE on the study of bright XRBs and bright AGNs is very significant, and the combined effect of spectroscopy of bright objects using Chandra, XMM-Newton and Suzaku was enormous. Of course, the deep imaging capability of Chandra had a tremendous impact on almost all branches of astrophysics: AGN feedback \citep{Fabian2012} and outflows \citep{KingPounds2015}, Ultra-luminous X-ray sources \citep{Kaaret2017}, dark matter haloes \citep{WechslerJeremy2018}.
 
 Here, I will subjectively divide the complexities of X-ray satellites (XS) as improved sophisticated versions which made a quantum leap from the previous versions, like XS v1.0 (UHURU type), XS v2.0 (HEAO-A/B  type) and XS v3.0 (Chandra/ XMM-Newton type). Are there any attempts this century to go for the next-generation satellites (XS v4.0)? Yes, attempts like the LOFT satellite with an order of magnitude larger area than RXTE and Athena with an order of magnitude larger area than Chandra were proposed, but we are still waiting to see a launch. With its bolometer, ASTRO-H, or Hitomi (2016, 2700 kg), could be considered as XS v4.0 for its spectroscopic capabilities, but unfortunately, it lasted only a month (ASTRO-H2, or XRISM - 2023, 2300 kg - was launched recently).

\begin{table}[htb]
\caption{Missions with specific goals in this century }\label{Mission00} 
\begin{tabular}{lccl}
\hline
Name& Operation & Mass (kg)  &Instruments \\\hline
Swift&2004 - & 1470& Burst Alert Telescope - BAT  (5200 cm$^2$)\\
& & & $~~~$ 15 - 150 keV\\
& & & X-ray Telescope- XRT (135 cm$^2$)  \\
& & & $~~~$ 4"; 0.2 - 10 keV\\
& & & UV Optical Telescope - UVOT \\
& & & $~~~$ 0.9''; 30 cm dia mirror; \\
NuSTAR &2012- & 350 & Hard X-ray focusing telescope\\
& & & $~~~$1'; 60 cm$^2$ at 78 keV; 3 - 78  keV    \\
IXPE &2021- & 330 &X-ray polarimeter (1 - 8 keV) \\
& & & $~~~$0.5"; 500 cm$^2$ at 2.3 keV   \\
\hline
\end{tabular}
\end{table}

\begin{table}[htb]
\caption{AstroSat and HMXT}\label{MissionAstro} 
\begin{tabular}{lccl}
\hline
Name& Operation & Mass (kg)  &Instruments \\\hline
AstroSAT&2015 - & 1513 &  Large Area X-ray Proportional Counter - LAXPC (6500 cm$^2$)\\
& & & $~~~$ 3 - 80 keV\\
& & & Scanning Sky Monitor - SSM (60 cm$^2$)  \\
& & & $~~~$  2.5 - 10  keV\\
& & & Soft X-ray  Telescope (SXT) (90 cm$^2$ at 1.5 keV) \\
& & & $~~~$  0.3 - 8  keV\\
& & & Cadmium Zinc Telluride Imager - CZTI (1000 cm$^2$) \\
& & & $~~~$ 20 - 200 keV\\
& & & UV Imaging Telescope - UVIT  \\
& & & $~~~$ 1.8''; 38 cm dia mirror \\
HXMT&2017- & 2800&Hard X-ray Modulation Telescope\\
& & &  Low Energy Telescope  (384 cm$^2$) \\
& & & $~~~$ 0.7 - 15  keV\\
& & &  Medium Energy Telescope (952 cm$^2$)\\
& & & $~~~$  5 - 30  keV\\
& & &  High Energy Telescope (5100 cm$^2$)\\
& & & $~~~$  20 - 200  keV\\
\hline
\end{tabular}
\end{table}

 On the other hand, this century can be considered the era of small-scale niche experiments in X-ray astronomy. These could be XS v.2.0 type but concentrated on specific topical issues. Three satellites of this class are noteworthy (Table  3). Swift (2004, 1470 kg), though a satellite meant for GRB detection, localisation, and afterglow studies, can swiftly move the satellite to any desired direction. This enabled the focussing telescopes (XRT and UVOT) to be used as X-ray and UV  monitors. The hard X-ray detector, BAT, had a considerable area (5200 cm$^2$) and provided the deepest hard X-ray sky survey. NuSTAR (2012, 350 kg) extended the focussing advantage to hard X-rays. The recently launched IXPE (2021, 330 kg) finally realised the potential of X-ray polarisation studies.

 \subsection{AstroSat: the major Indian contribution}

\begin{figure}[h]
\centering
\includegraphics[angle=-90,width=\textwidth, trim={2cm 3cm 6cm 3cm},clip]{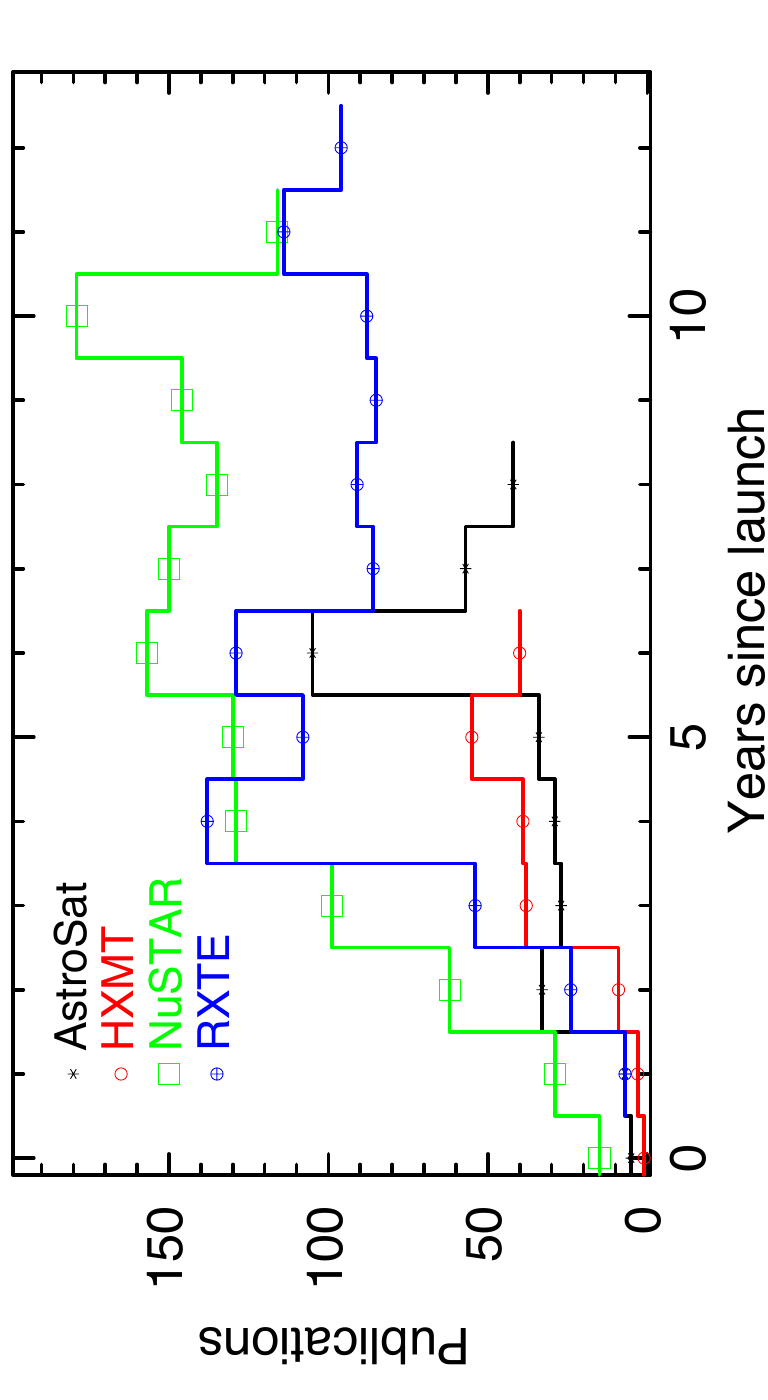} 
\caption{Number of refereed publications based on data from  currently operating hard X-ray instruments NuSTAR (launched in 2012), AstroSat (2015) and HXMT (2017), compared to those based on RXTE (1995) data.}
\label{fig2}
\end{figure}

The central story of Indian effort in this era is the successful launch and operation of the AstroSat satellite.
 
 AstroSat (2015, 1500 kg)  can be considered a second generation X-ray satellite (XS v.2.0) with two complementary science priorities: a wide-angle UV telescope with excellent seeing (UVIT) and a suite of co-aligned X-ray telescopes covering an extensive energy range of 0.3 to 190 keV. Along with NuSTAR and HXMT (2017, 2800 kg), AstroSat data have the potential to make inroads into the long-neglected hard X-ray astronomy (Table 4). 
 
 The story of the making of AstroSat is told in Paper1. Here, I will discuss the sociological aspects. The fortuitous combination of multiple factors positively reinforcing each other for  IXAE development saw an improved version for AstroSat. The same factors are discussed below at some length.

This century saw the emergence of many research centres in India practising X-ray astronomy apart from the traditional strong bases at TIFR and ISRO: Indian Center for Space Science (ICSP) at Kolkata, Inter-University Center for Astronomy and Astrophysics (IUCAA) at Pune, Indian Institute of Astrophysics (IIA) Bengaluru, Physical Research Laboratory (PRL) Ahmedabad, Raman Research Institute (RRI) Bengaluru and many Universities and educational Institutes. The easy availability of RXTE data, coupled with an effort to hold regular training workshops, built up a reasonable number of X-ray data users and interpreters. The effective number of core X-ray astronomers in India, that is, X-ray astronomers who could build instruments, develop X-ray data analysis codes, or manage X-ray missions, however, remained a handful (say, 5 to 10) throughout the past six decades of X-ray astronomy. Though it is creditworthy that a small number of X-ray astronomers could dedicate a lifetime to realise AstroSat, several negative aspects of this approach prevented the healthy growth of X-ray astronomy in India. 

Compared to the IXAE effort, the technical support for AstroSat by the pool of technical expertise from the various ISRO centres was of vastly superior nature and content.
While discussing the growth of creative talent for X-ray astronomy, I emphasised that the growth of talent by collective, collaborative efforts could be as significant, if not more, as the selection of good talent. In hindsight, I can list several essential ingredients that could have been responsible for developing collective creative talent that has emerged at ISRO and extensively used by X-ray astronomers to realise AstroSat. I am pursuing my core belief that X-ray astronomy is a creative art and, even more importantly, a collective creative art. I am examining the events through the prism of understanding the talent growth for such a creative art.

\begin{itemize}
\item
{\bf Long lead time for talent growth:}
TIFR and ISRO, though both traced the development of Science and Technology in post-independent India, grew with their distinctive styles. TIFR concentrated on making extra effort to identify and encourage talents of high order. ISRO, on the other hand, though led by talented individuals at various stages, required a considerable pool of dedicated individuals, quite often for doing and developing mundane things that were not highly challenging, intellectually, but nevertheless required dedicated and continuous efforts.  The fact that they were assured of continuous appointments further helped develop pools of expertise at ISRO.
\item 
{\bf Specific projects:}
Since India was under a Western embargo for smooth technical transfer of knowledge for various reasons during the past six decades, several technical areas of expertise for space sciences had to be developed in India from the first principles. Though most of these technologies existed elsewhere, discovering new ways of doing things independently has its creative charm. Additionally, these technologies had to be made operational for specific purposes of launching or operating a satellite, thus leading to a sense of purpose and completion for developing the know-how.
\item 
{\bf Attitude of getting things done:} 
Launching and operating a complex satellite requires both a culture of science and a culture of making things work, that is, engineering. The engineering challenges are very specific and product-driven, which helps develop a culture of mutual help and collective work. Coming from TIFR, which had a culture of extreme individualism, I found it refreshing to see a culture of collective creative effort in ISRO.
\item
{\bf Freedom with control:}
Unlike a private organisation where strict control of a worker's productivity and expertise can be monitored, nurtured or optimised, ISRO essentially is a government organisation. Hence, extreme control of an individual worker was not entirely possible. The team leader's job was to motivate and get the work done, leaving sufficient leeway for the workers to innovate and develop workable solutions to the multiple tasks. It was quite pleasing to see that ISRO had developed a culture of balanced control and freedom to get the best out of a team for specific targeted projects. Though this method did have certain negative connotations, like the tendency to isolate highly talented individuals but having strong, uncompromising individualistic stands, the overall effect was quite positive.
\end{itemize}

 The above is a subjective list of the probable causes for the emergence of significant talent pools at the various ISRO centres that were extensively used to develop AstroSat. All instruments were successfully built and operated. They performed (and are still performing) as designed. The limitations were mainly two-fold. Firstly, the number of experienced scientists available in India is limited. This led to a delay in the realisation of AstroSat and, more importantly, in my opinion, made the scientists indulge in diverse necessary activities, thus preventing the emergence of deep expertise in any specific area of research. The culture of ISRO, on the other hand, is excellent in tackling some specific tasks in a project mode in a specified time but not entirely amenable to the realisation of some long-standing open problems. The primary developmental activities of AstroSat happened a few years before the launch of AstroSat in 2015: after some visibility in the realisation of the instruments and after the Indian science management decided to fully endorse AstroSat and work towards it. The work before this intense project mode involved some individual scientists toiling towards the instrument's realisation and some help from specific ISRO laboratories. After the launch, the work reverted to this mode of `few scientists-and-few-centres' style, severely affecting the effective use and utilisation of AstroSat data.

\subsection{A critical summary of the third era of X-ray astronomy}

\subsubsection{The science context}

If I were to revisit the speculation from an early review that  X-ray astronomy could be `..one of the most fruitful channels of data about the largest scales of the universe', or it could be relegated  `to the detailed study of certain unusual stars' \citep{Morrison1967}, what would be my take?

 Mostly the latter.
 
     Here, I highlight the factors limiting the impact of X-ray astronomy on our broader understanding of the universe. 
   
   Among the three key research areas highlighted earlier, viz., the study of X-ray binaries, X-ray spectroscopy, and deep imaging,  the third one is directly relevant to the quest for the `largest scales of the universe'.   Deep X-ray imaging is possible using large area focussing X-ray detectors like Chandra. Due to the peculiar nature of grazing incidence mirror technology, the effective area of the X-ray mirrors drops very sharply with energy; hence, the imaging is done mainly at low energies (1 - 2 keV). Only the thermal part of the X-ray spectra are probed at these energies. Though the temperature ranges probed by X-ray observations are significantly higher than from observations using lower frequencies, to trace the thermal universe in the largest scales, the technical developments in other wavelengths (primarily optical and infrared) ensured a deeper reach into the far universe (particularly after the launch of JWST). These developments have relegated X-ray imaging to a supporting role. There exist, though, some very niche areas, like the study of the temperature distribution in clusters of galaxies, where X-ray observations give a unique perspective  and, in recent times,  eROSITA (2019, 810 kg)  is making a mark in these areas. Mostly, however, the restricted improvement of the sensitivity of X-ray telescopes has not made X-ray astronomy the primary choice of wavelength to study the  `largest scales of the universe'.

  The first two areas, though, pertain primarily to `the detailed study of certain unusual stars', a detailed characterisation of the exotic non-thermal phenomena has far-reaching implications for our understanding of the largest scales of the universe. If the X-ray detectors, particularly in the hard X-ray range (more on this in the next section), can clearly decipher the source emission mechanisms to unambiguously elicit pertinent information on the emitting region like the source geometry, energy source and the details of the emission mechanisms,  it will have tremendous impact on our understanding of the largest scales of the universe by, for example, using gamma-ray bursts as distant indicators or Active Galactic Nuclei to understand the growth of black holes. Though NuSTAR and Swift have provided some important inputs, and IXPE is getting some exciting results on X-ray polarimetry, I would rather say that there were no groundbreaking results in our understanding of, say, black hole sources and their environments. Black hole masses are measured by velocity measurements through optical spectroscopy, and the spin measurements by X-ray observations are far from convincing
  \citep[see][for a detailed discussion on the limitations of X-ray sensitivity for our understanding of the accretion phenomenon near black holes]{Rao2024}.
  
  In the next section, I will delve into the sensitivity improvements in X-ray measurements. This will help us understand why X-ray astronomy has yet to make deep inroads into the non-thermal phenomena in cosmic sources.

   \subsubsection{Sensitivity of X-ray detectors}

 Let us understand the spectroscopic capabilities of X-ray telescope/ detector systems for the first two objectives: spectral and timing studies of bright XRBs and continuum spectroscopy of a set of reasonably bright sources. The former are bright objects up to the brightness of Crab, and the latter are mostly a few milliCrab sources. 
   
     I have taken the data for Crab itself to examine the current state of the  X-ray spectral sensitivity for Crab-like sources, particularly at higher energies ($>$ 50 keV). NuSTAR is the only operating hard X-ray focussing instrument, and I have taken the Crab data based on an observation made on 2016 May 3 (see Table 5). Many hard X-ray collimated detectors are currently operating, and they all have the effective area of the order of $\sim$ 1000 cm$^2$ and hence are of similar sensitivity. Background estimation is an intricate task at these energies, and it is imperative to make an accurate background measurement to get good spectral sensitivity using collimated detectors at higher energies. CZT Imager (CZTI) of AstroSat \citep{Bhalerao2017} uses a box type Coded Aperture Mask (CAM) arrangement in a collimator with a medium-sized field of view  (4$^\circ$.6 $\times$ 4$^\circ$.6), enabling simultaneous background measurement \citep{Mithun2024}. I have taken the Crab data obtained from CZTI observations in the 20 -- 200 keV range. The NuSTAR and CZTI data are fit simultaneously with a power law, and the unfolded energy spectra are shown in the top part of Fig 3.
     
    As an example of a bright AGN, I have taken the simultaneous observations of IC 4329A made using XMM-Newton and RXTE on 2003 Aug 3, and the spectral data from RXTE-PCA and XMM-Newton EPIC-pn are simultaneously fit using a model consisting of a power law and reflection component and Iron lines (model 3 in  Markowitz et al. 2006). The unfolded energy spectra are shown in the bottom part of Figure 3. Note that a 25\% area renormalisation is required for RXTE-PCA compared to EPIC-pn. NuSTAR also observed IC 4329A on 2012 Aug 12. The details of all these observations are given in Table 5. To have a uniform spectral binning, all data are logarithmically re-binned to have 50 spectral bins in a decade of energy.
    
      The unfolded energy spectra as a fraction of Crab flux are given in Figure 4. In the same figure, the sensitivity of each instrument, deemed as the errors in the data for the uniform logarithmic binning of 50 data points per decade, is also shown. It can be seen from Table 5 that the durations of observation are vastly different for each instrument. Hence, the sensitivity numbers are rescaled as the square root of time in the units of 10 ks, as shown in Figure 5. Also included in this figure are the data for IC 4329 A using NuSTAR observations made in 2012. Note that the spectral sensitivity of NuSTAR for bright sources like Crab is a few factors inferior compared to the AGN observations because NuSTAR has the limitation of storing only about 300 photons per second, thus limiting the effective observation time for bright sources \citep{Harrison2013}. 
      
       Several aspects of the sensitivity of X-ray instruments can be discerned from Figure 5.
     
\begin{itemize}

\item{The huge improvement in sensitivity by focussing techniques is mainly confined to very low energies. At around 1 keV, XMM-Newton EPIC-pn has a few $\mu$Crab spectral sensitivity.}

\item{The reflectivity of grazing incidence X-ray telescopes drops very sharply with energy and beyond 6 keV, and in the energy region of 10 - 20 keV, the spectral sensitivity of focussing X-ray instruments as well as the large area collimated detectors are modest and comparable to each other. The sensitivities are in the range of 0.1 to 1 milliCrab. }

\item{Above 20 keV, the sensitivity drops sharply with energy due to the sharply decreasing effective area for X-ray mirror systems and sharply increasing background for collimated detectors. }

\item{The sensitivity is quite poor in the hard X-ray regime of  50 - 200 keV.}

\end{itemize}

It is doubtful that there will be drastic improvements in the spectral sensitivity of X-ray instruments in the future. The area drops drastically with energy for X-ray mirrors, and above 70 keV, it is very, very challenging to build them. For the collimated detectors, a high particle-induced background is a serious issue. Hence, we may have stupendous sensitivity below 6 keV for continuum X-ray spectroscopy but medium sensitivity in the 10 - 50 keV region and poor sensitivity above $\sim$ 50 keV.

\begin{figure}[h]
\centering
\includegraphics[angle=-90, scale=0.5,trim={1cm 2cm 2cm 2cm},clip]{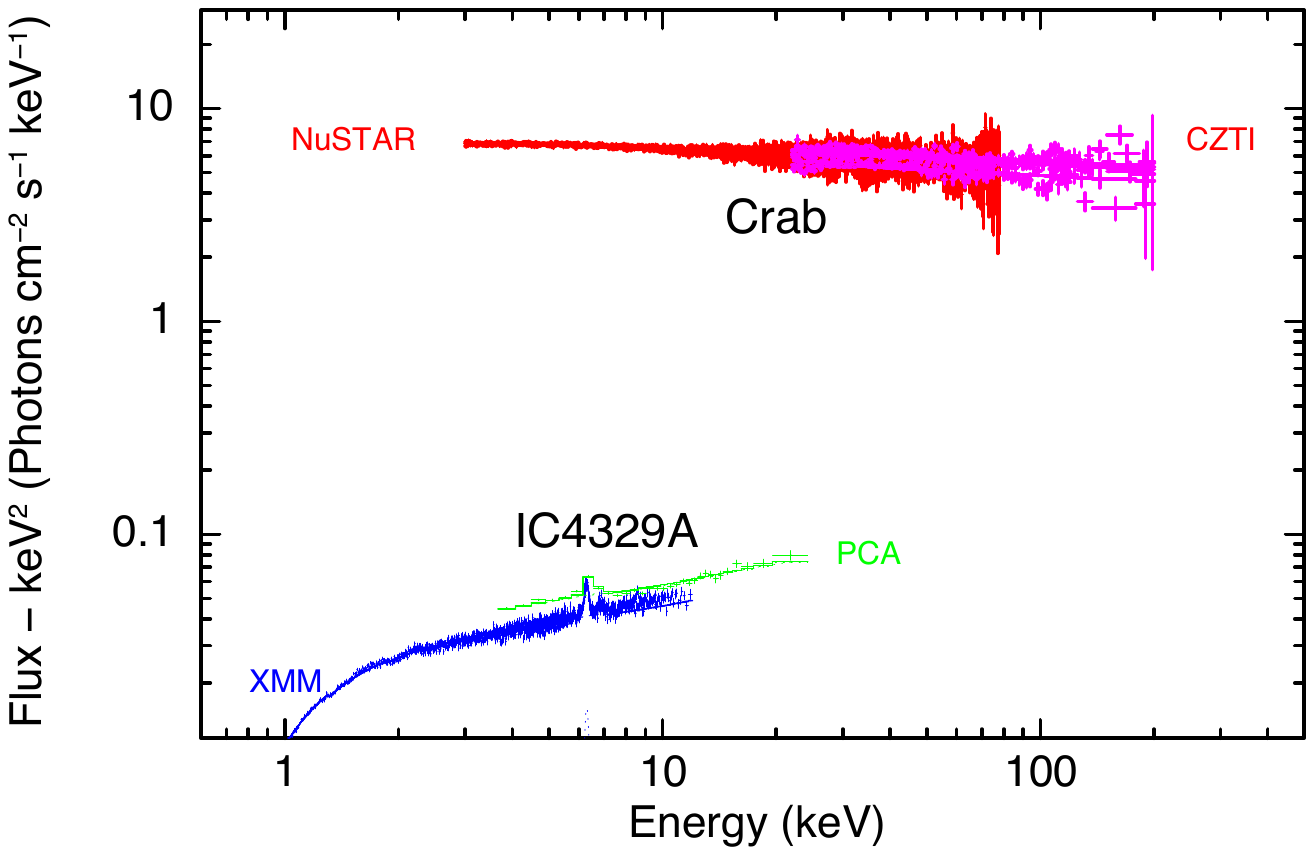} 
\caption{Good quality X-ray spectra of representative X-ray sources: Crab (top part of the figure), as measured by NuSTAR and AstroSat-CZTI and a bright Seyfert galaxy IC 4329A (bottom part of the figure), as measured by XMM-Newton-EPIC-pn and RXTE-PCA. }
\label{fig3}
\end{figure}

\begin{table}[htb]
\caption{Data used for spectral fitting}\label{MissionAstro} 
\begin{tabular}{lllccr}
\hline
Mission& Instrument & Source & Date & Time & Duration (ks) \\\hline
XMM-Newton&EPIC-pn & IC 4329A & 2003 Aug 3 & 06:56:27 & 136.0 \\
RXTE & PCA & IC 4329A & 2003 Aug 3 & 09:32:16 & 19.3 \\
NuSTAR & FPM A\&B &  IC 4329A & 2012 Aug 12&16:06:07  &162.4\\
NuSTAR & FPM A\&B & Crab &2016 May 3 & 19:10:12& 5.2 \\
AstroSAT&CZTI &  Crab & 2016 Mar 31 & 05:39:19 & 226.9\\
\hline
\end{tabular}
\end{table}

\begin{figure}[h]
\centering
\includegraphics[angle=-90,scale=0.5,trim={2cm 2cm 2cm 2cm},clip]{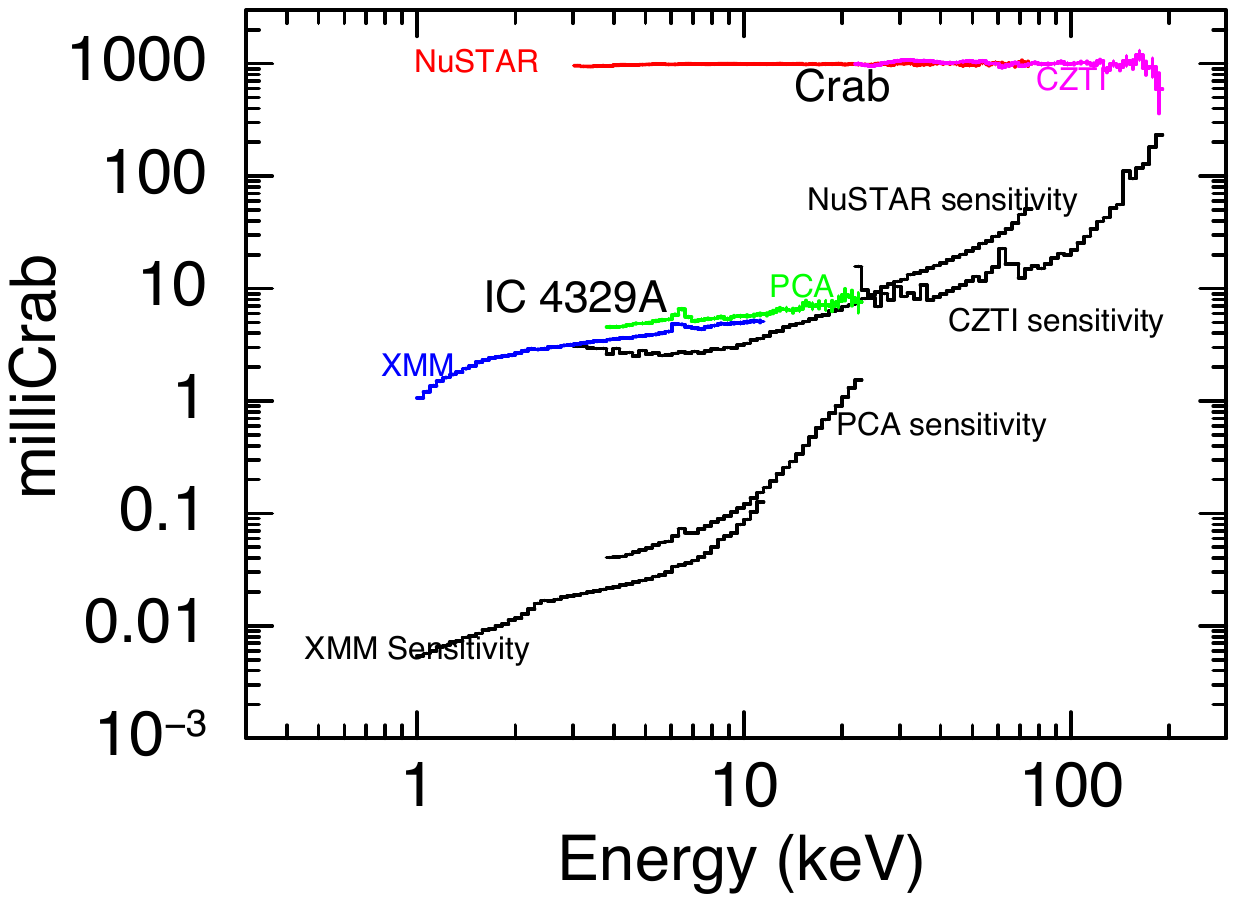} 
\caption{Unfolded spectra of Crab (data from NuSTAR and CZTI) and IC4329A (data from XMM-Newton and PCA), in the units of milliCrab and shown in colors. The errors in the data points are deemed spectral sensitivity and plotted in the figure. CZTI appears more sensitive than NuSTAR due to the higher exposure; see Fig. 5 for an exposure-corrected sensitivity plot.}
\label{fig4}
\end{figure}

\begin{figure}[h]
\centering
\includegraphics[angle=-90,scale=0.5,trim={2cm 2cm 2cm 2cm},clip]{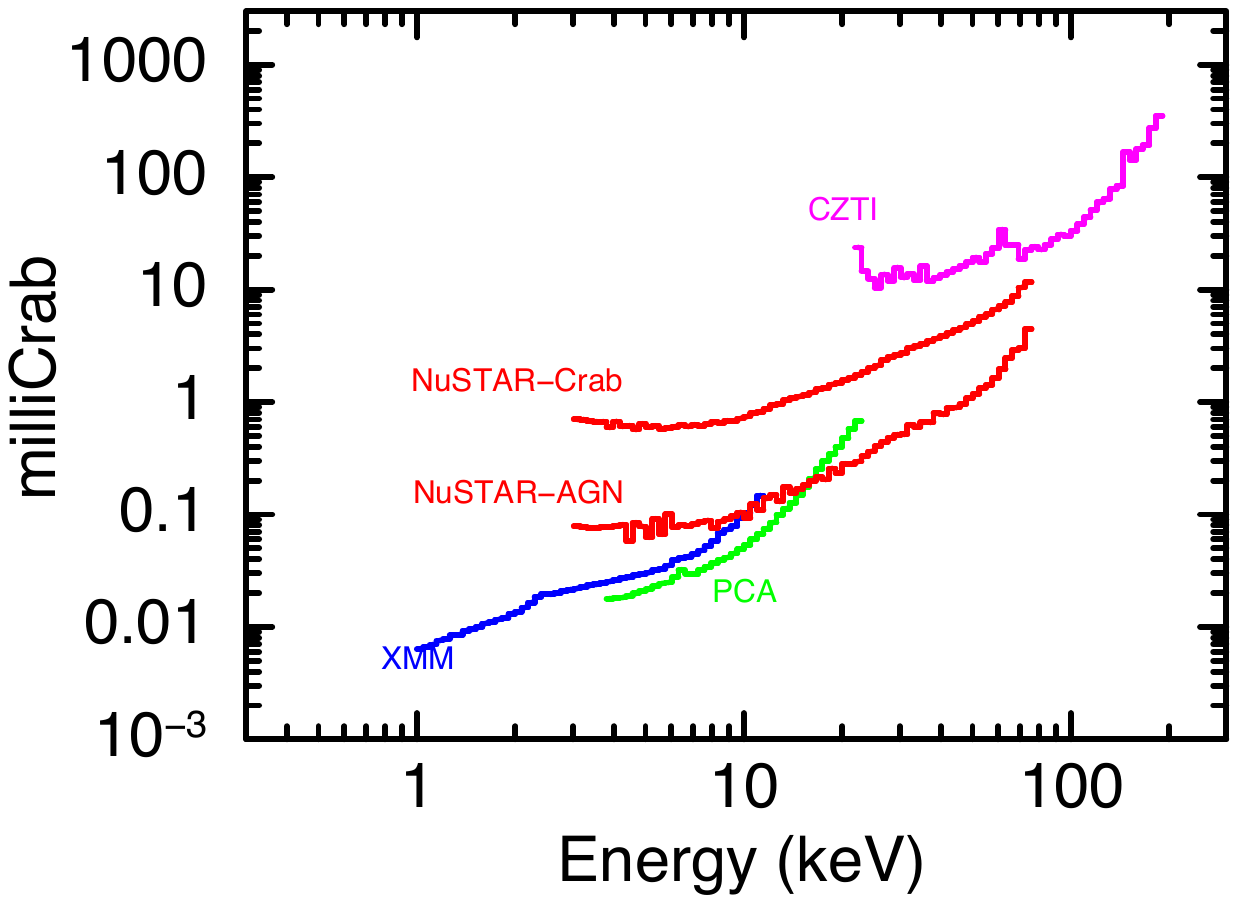} 
\caption{The sensitivity values (from Figure 4) are plotted here after normalising them to a uniform observation time of 10 ks. Also plotted is the sensitivity of NuSTAR instrument for the observation of the AGN, IC4329A. An earlier version of this figure is also given in \citet{Rao2024}.}
\label{fig5}
\end{figure}

\subsubsection{Impact on science due to the poor sensitivity of X-ray detectors above 50 keV}

As mentioned earlier, X-ray astronomy straddles the thermal and non-thermal astrophysical regimes. 

Here, I want to highlight the phenomenal differences in how these two branches of astrophysics are done.

When we peer into the cosmos, we observe a vast variety of exotic phenomena. Most of these can generally be understood by the known laws of Physics. Hence, it appears that the primary task of an astronomer is to carefully make observations, note down possible boundary conditions, and diligently apply the known laws of physics to explain the observations. If a new observation indicates contradictions to the previously worked-out solutions, it is generally surmised that either the assumed boundary conditions are wrong or the previous observations had limitations. It is only a question of further diligent work to iron out these contradictions. 

The above approach is exceptionally well suited to study any astrophysical phenomena that have had sufficient time to stabilise and thermalise, particularly those objects that change in years or decades of time scales. For one thing, the variable `time' can be taken out of the considerations, and one can go back and observe the object or phenomena with better equipment to refine the measurables. 
Any slowly time-varying phenomena, too, can be understood with the same approach by studying the systems as a series of time-stamped pictures.

 Hence, the crux of better understanding is better measurements.

 However, we do observe many phenomena in the universe which can be called non-thermal: the energy loss rate is comparable to or faster than the energy generation rate. As a practical example, consider accretion phenomena near stellar mass black holes. The typical time scale of energy generation is in milliseconds. Since the energy source is gravitation and the gravitation potential goes as the reciprocal of distance from the black hole, most energy is released very close to the black hole, and most time scales (like the free-fall time scale) are in the millisecond range. Further, the energy gain is up to a significant fraction of the rest mass energy and thus, for neutral materials, the electrons gain an enormous amount of energy (equivalent to close to a thousand times their rest mass energy), and they copiously emit radiation with energy peaking in hard X-rays, the very branch of electromagnetic radiation where the observing efficiency is very very low. Another intricacy is the sort of Physics governing these phenomena. To start with, we are not entirely sure that event horizons exist.   The intricacies of the energetics in an environment never probed in the terrestrial environment (high gravity, high magnetic field) add further layers of complications.
 
  Hence, the methods commonly employed to understand the thermal phenomena in the universe may not apply to understanding the non-thermal phenomena. The time scales are too fast, observations are meagre, and the Physics principles are uncertain. 
 
 The third era of X-ray astronomy failed to make inroads into this very difficult but extremely important branch of astrophysics. In the next section, I will dwell on the sociological aspects that could have influenced this lack of progress in non-thermal astrophysics.

  \subsubsection{The way of doing science...} 
 
 The way of doing science has changed during the past few decades.
 
 One of the most significant changes in the way of doing science in the late twentieth century, as compared to the nineteenth and early twentieth centuries, is the dramatic increase in the number of people doing science. Instead of being an exclusive club of talented individuals pursuing their passion, science moved onto the universities which received massive funding, particularly in the USA and Europe. Science was firmly coupled with education.
 
 What are the consequences?
 
  While discussing the effect of public science on corporate R\&D, \citet{Arora2023} conclude that, as far as the effect on industry goes, the university-led R\&D effort did not significantly impact industrial productivity.
   In the realm of astrophysics, concerns are expressed about the way science is done. \citet{Antonucci2013}, while reviewing our understanding of Quasars, points out that `Quasars still defy
explanation' and `...having given up on understanding AGNs, the community now focuses on the more modest goal of counting them'. I have also noted earlier that in recent times, there is   `...a lack of progress in the long term and deep thoughts' in astronomy and astrophysics
\citep{RaoCurrentSci}. Following \citet{Antonucci2013}
 I might also add that `having given up on understanding accretion onto black holes, the community now focuses on the more modest goal of categorising QPOs'.

 I want to examine the implications of the fact that the number of people doing science has increased tremendously and offer a possible explanation for the feeling that deeper science problems, in recent times, are not addressed.  
  Further, I would like to explore the possibility that the current mode of doing science, which relies heavily on university teachers and corporate-style funding, is not conducive to tackling complex, open-ended problems.

  There are several obvious aspects of the current style of doing research that are not very conducive to tackling complex  problems. 
  Since practising scientists have many additional responsibilities (teaching, administration, etc.), the incentive is always to do quick publishable research. Added to this, the founding method is tuned to encourage predictable deliverables. Hence, the incentive to take up complex problems with uncertain outcomes is quite low.
  
  There could be, however, one more additional prominent reason for the failure of the scientific community to take up extremely challenging problems. 
        
 Science is a creative art and requires talent to deliver. Though various types of screening are done to select the most talented people at the various stages of entry levels in science, it is highly important that the practice of science also emphasises talent growth. The obvious way of growing talent is through the continuous and conscious practice of the trade, particularly for collective creative talents like X-ray astronomy, and also to interact deeply with other talented people.
The current scenario of 
providing enormous incentives for short-term products is possibly harming the requirement for talent growth.

$~$

{\bf Hence, I want to advance the hypothesis that stunted talent growth is the primary reason for the inability of the community to take up challenging problems.}

Let me re-articulate, re-emphasise and expand on this point.

$~$
The input stream of scientists was motivated by the individual romantic fascination with astronomy and also as a profession. Since it was assumed that science is important to society (as a unique style of thinking - the so-called scientific temper - and also to deliver scientific products), significant investments were made to make science a profession monetarily attractive. A large number of `entry-level' scientists were trained at graduate schools, mostly coupled to the universities. Some are eventually selected from this pool as professional astronomers, primarily in the university sector. 

A professional X-ray astronomer is required to develop better and better instruments to make good observations and then interpret these observations based on the known laws of physics.  
Since the amount of work needed to implement and operate a space science experiment successfully is quite huge, several characteristics have evolved into the system. I will highlight some of them which could be relevant to the theme of reduced talent growth in the community. Firstly, a methodology was evolved to break down the task into minute parts so that the work could be distributed to many people, including those with very low training or capability. This, in effect, increased the management burden on established X-ray astronomers. Secondly, a large number of these foot soldiers are young students in the university setup, thus making teaching and training these next-generation scientists a necessary activity of the established scientists. This took away a substantial chunk of their time. Lastly, since most activities are established and routine, the efforts that go into actual creative activities are few and far between. Though some individuals could retain their creative instincts till late in their career, the deluge of work resulting in the explosive growth of the number of publications (even the worst papers need to be refereed) generally stifles good collective creative work.

 The above is a cartoonish sketch of the current style of working in X-ray astronomy, but, nevertheless, it provides a possible indicator of the lack of deep thought in most of the recent scientific activities in X-ray astronomy.

In this perspective, the difficulties of doing hard X-ray astronomy and the consequent limited advance in non-thermal astrophysics are symbolic of the limitations of the way of doing science that is currently practised. The lack of progress in hard X-ray astronomy and the understanding of the non-thermal universe is merely, I venture to add, a byproduct of this style of doing research.

\subsubsection{The Indian scenario}

In this century, the work on X-ray astronomy in India diversified to many centres other than the traditional strongholds like TIFR and ISRO centres. Instrumentation work was undertaken at PRL, RRI, and ICSP. ICSP also led a very strong theoretical and observational effort in X-ray astronomy. RRI was deeply involved in the development of a dedicated X-ray polarisation satellite.

The significant advance in X-ray astronomy from India in this era, however,  is the launch and operation of the AstroSat satellite.

 AstroSat, fortuitously, has all the ingredients to make a mark in the difficult field of hard X-ray measurements. 
 
   The large area X-ray detectors of LAXPC can probe X-ray variabilities in the millisecond range to understand the dynamics of accretion onto black holes for bright Galactic X-ray binaries. LAXPC detectors have the highest effective area among all X-ray detectors flown so far in the 30 -- 70 keV range, thus providing crucial timing information in this critical energy range. These data are beautifully complemented at lower energies with SXT down to 0.3 keV, with very good energy resolution. All attempts to use RXTE PCA data along with good energy resolution low energy detectors like those in Chandra and XMM-Newton had only limited success. For one thing, Galactic X-ray binaries are too bright for these sensitive soft X-ray detectors. Secondly, coordinating simultaneous observations using different satellites is quite difficult. SXT, by design, can deal with the pulse pile-up effect from bright sources, and it is in a platform meant for simultaneous observations with LAXPC.

Similarly, getting simultaneous hard X-ray measurements at higher energies was challenging. CZTI of AstroSat is equipped with a box-type collimator that enables simultaneous background measurement up to 200 keV, with an additional input of polarisation measurements for bright sources. There are also other features in AstroSat that are very useful for making wide-band X-ray spectral measurements. 
 The satellite is in a near-equatorial low earth orbit, just skimming the dreaded high background South Atlantic Anomaly (SAA) region and thus providing a stable background. All the instruments transmit time-tagged data for individual photons. This is very useful for understanding possible systematic uncertainties: one can examine the data at all time scales and use the Poisson nature of individual photon interaction to address, understand, and correct systematic errors in the data. AstroSat also has an X-ray all-sky monitor, SSM, which followed closely the design of the highly successful ASM of RXTE, but has several interesting additional features like having time-tagged data for individual photons, low background due to the equatorial orbit of AstroSat and the ability to stare at any interesting object for longer durations. 
 
 AstroSat has been working for over eight years, and the instruments performed as expected. The number of refereed publications compares well with other similar observatories (see Figure 2).
   
   A closer inspection of the publication record from AstroSat reveals the following. 
  
   As of 2024 January, there are a total of 1435 publications with the word `AstroSat' in the abstract, as per the ADS records (there indeed are some papers based on AstroSat data without explicitly mentioning the word `AstroSat' in their abstracts, but these are a few and should not dilute the broad conclusions we are trying to draw). A significant fraction of these publications are in GCN circulars (678).
   The number of refereed publications is 413, 14 in high-impact journals like Nature, Nature Astronomy and ApJ (Letters). Two of the prime instruments in AstroSat follow the legacy of previously highly
   successful observatories: LAXPC is an improved version of RXTE PCA, and UVIT is that of GALEX. The RXTE PCA heritage of using timing data to examine the spectral and timing properties of bright X-ray binaries  
   is quite helpful for entry-level scientists to get into the subject. Similarly, entry-level scientists can also use the wide FoV images from GALEX. Many university students thus benefited from following these traditions in analysing LAXPC and UVIT data. 
   
   Hence, to understand the trend of the most important results from AstroSat, I have examined the highly cited papers from AstroSat.
Among  the 51 most cited papers (citations more than 20) 
23 are instrument papers or review papers.
Out of the remaining 28, 
    12 pertain to X-ray timing of bright sources (mainly LAXPC driven) and 
     9 from UVIT (stars/ galaxies/ clusters) and 
     7 from CZTI (GRBs/ polarimetry).     
Most of the results are authored by core instrument team members.
Further, there are 
very few results from multiple instruments in the list of highly cited papers.

Thus, the promise of AstroSat to make substantial headway into the difficult arena of non-thermal astrophysics is yet to be realised. A methodology to estimate and correct the background using all background indicators from all X-ray instruments is yet to emerge. A rigorous attempt to pit the individual instrument responses with each other and instruments in other observatories using simultaneous and quasi-simultaneous observations of multiple sources and arriving at an authoritative understanding of the systematics in the responses is currently a work in progress. The promise of SSM as an improved monitor compared to ASM (of RXTE) is indeed very far from realisation. One can say that a huge amount of rigorous work remains to be done, for each of the instruments individually as well as jointly,  to carefully mine the extremely important wide band X-ray data from AstroSat. Also, the operation and analysis are not nimble enough and  the analysis method has not yet reached the sophistication of the many of the established observatories, leading to an under-utilisation of data.


Though it is very credible that a second generation X-ray satellite (XS V2.0)  was conceived and launched from India, and much effort has gone into the operation of the satellite and smoothening data analysis,  the under-utilisation of the precious AstroSat data perhaps points towards a larger malaise.
 Now, I want to tie this up with my observations about how science is done worldwide. The negative effects I outlined earlier had an exaggerated impact in the Indian context. The quality of entry-level scientists is relatively lower, the number of established scientists is few, and the burden of multiple tasks like management, administration and teaching is amplified.  I am advocating that the style of doing science does not encourage talent growth; this can have an exaggerated impact in the context of Indian science.

 

 Looking back at the development of X-ray astronomy in India over the past six decades, we can treat these developments as a sort of experiment at talent growth. The idea of TIFR, to identify potential talents and give them full encouragement, resulted in a set of experts in X-ray astronomy, able to do meaningful experiments but fell short of doing truly impactful research, my interpretation of this being their inability to work together and in cohesion. 
 The spread of X-ray astronomy in this century mostly followed the Western model of coupling research with education. Hence, the multitude of tasks required to be done is quite enormous. Further, the X-ray astronomy community is still tiny, thus enhancing the necessary mundane work for each researcher, further limiting the growth of talent.
 
   As per my observations, the most worrying aspect is the lack of growth as researchers. Hence, many of the routine tasks of AstroSat got done, leaving much to be desired in the intricate tasks of going into depth 
   the finer aspects of the work like cross-calibration, response building, etc.


\section{Summary and conclusions}

 This article started with an ambitious and vast canvas: understanding the evolution of X-ray astronomy during the past six decades, juxtapositioning it with Indian developments and deriving some meaningful lessons. By the very nature of the topic, in several places, the article's tone was more general and perhaps lacking in rigour, but I have tried to distinguish clearly between information and surmises.  
 
 X-ray astronomy is a branch of space astronomy, an area of science requiring very high resource investment. The national pride of being the first to achieve some extraordinary feat was the driving force behind the huge investments, particularly in the Apollo era of the sixties. This led to very rapid developments in X-ray astronomy, opening up this new window in the electromagnetic spectrum to observe and understand the universe. X-ray astronomy enriched the wisdom already gained by the use of the explosive growth in physics coupled with optical and radio observations in the early and mid-twentieth century, and it also 
 enabled us to peep into some exotic observations like emissions from close to black holes. The peace and accelerated growth of wealth in the developed countries in the post-Second World War era ensured further ploughing of resources into space sciences, including X-ray astronomy,  enabling it to grow as a mature field. 
 
 Though X-ray astronomy is contributing meaningfully to our quest to understanding the cosmos, it is argued that there are niche areas where X-ray observations, particularly in the higher energies, can, in principle, help obtain very insightful answers to several critical questions in astrophysics like the nature of black holes, growth of black holes, and disk-jet connection in accreting objects. Two obstacles are identified. It is demonstrated that the improvement in the sensitivity of observations in the higher energy X-rays is not very significant, and it is unlikely to improve in the near future, thus limiting the effectiveness of a critical observational area. The second obstacle is more sociological. The nature of X-ray astronomy demands the execution of complex projects in a corporate style requiring clear-cut targets and deliverables, thus sidelining any area which requires a nebulous and exploratory style of research. As an off-shoot of this style (along with the effect of coupling research with education), it is proposed that continuous talent growth in researchers is hindered, thus preventing a collective and deep investigation of complex and subtle problems.  
 
 The developments in India grew in parallel with the trends of X-ray astronomy worldwide. It is creditworthy that with limited resources, India could initially replicate most of the developments in X-ray astronomy, particularly when the work was limited to balloon and rocket-borne experiments. It was, however, pointed out that in this game of catching up, no original or noteworthy results were obtained. Though lack of resources (material as well as human resources) could have played a major role, the inability to collectively address some deep problems was noticed. In recent times, an observatory class satellite AstroSat was launched and made operational, but extracting the most meaningful results from the AstroSat data, which could have made a deeper impact on the subject of non-thermal astrophysics, particularly in areas related to accretion onto stellar mass black holes, is a work that largely remained undone.
 
 It is hinted that the deluge of work that a professional X-ray astronomer has to face due to the job's complexity and the necessity of involving a huge number of entry-level scientists in the education sector could have inhibited robust and collective talent growth. This could be one of the core reasons for the lack of growth in hard X-ray astronomy (resulting in a lack of growth in our understanding of the non-thermal astrophysical phenomena) and the lack of more impactful results from AstroSat.
 
  Based on these insights,  I will give a personalised answer to some of the questions posed in the first section (Introduction).

\begin{itemize}
\item
{\bf Development of an area of science: individual or cultural?}\\
$~$\\
Based on the experiences of TIFR, where sustained efforts were made to support talented individuals, I would say that the development in a scientific research area is primarily cultural, though individuals with extra drive and talent play a critical role.   The individual's push and the societal fertility need to work symbiotically for a scientific area to flourish. From this perspective, the idea of TIFR, establishing a centre of excellence in isolation, unsurprisingly did not yield any great scientific results (in the context of X-ray astronomy). Still, nevertheless, it enabled the development of a fertile ground through which new emerging opportunities could be exploited (like the launch and operation of AstroSat). 
\item {\bf How did the corporatization of science, brought about by X-ray astronomy, influence the way of doing science?}\\
$~$\\
The most obvious positive effect is the quick realization of solutions to `known' problems. Once it was known that bright X-ray sources existed, it took less than a decade to send an X-ray satellite to study them in detail. Once it was known that the subsequent improvement in sensitivity necessarily required the development of X-ray mirrors, Einstein Observatory, with a good quality X-ray mirror, was up and working within the next decade. Once it was known that GRBs are cosmological and have X-ray afterglows, a swift-moving satellite, Swift, was launched in a few years, and redshifts of many GRBs were measured. The style of funding the expensive satellite programs ensured an adequate use of the data, or rather, the data from these instruments became the prime focus of the community's attention.

   Such a sledgehammer approach to problem-solving has the possible side effect of ignoring alternate ideas, dissents, open-minded questioning, and attention to subtle signals and thoughts. Though GRBs were detected in the sixties, their cosmological origin was identified three decades later with an almost serendipitous detection of the X-ray afterglows. The reasons could be that the methods to solve the problem of the origin of GRBs were not obvious, and the community's working style was not very conducive to exploring subtle and obscure signals.  
\item {\bf Whether there is still a place for blue-sky research?}\\
$~$\\
  There are two aspects to this question. Whether there should be societal support for blue-sky research and whether individuals should devote a disproportionately large amount of their time to doing blue-sky research. I am providing extremely speculative answers to these critical questions.
  
The intense levels of the articulation of the human mind, embodied in arts and science, have fascinated humanity for a long time, and there has always been societal support. I want to highlight some subtle differences over the period of time in this societal support to these individualistic activities. 

The world has seen explosive growth in wealth during the past century, and in recent times, a significant fraction of humanity has a comfortable life and needs not to worry about day-to-day survival; they can, in principle, indulge in their individual passions. This very significant change can have, and perhaps has, a subtle effect on the way society supports extreme talents.
In the earlier era, where society could not afford to support a large number of people indulging in specialised activities, unique talents could have been supported by a `queen-bee' style of grooming (that is, someone is selected quite early and given extra special treatment). Some signs of this style are a little arbitrariness in selecting individuals (luck too playing an important part), eulogising the selected ones, ascribing innate qualities and special reasons for their brilliance. 

  If, on the other hand, we accept my earlier hypothesis that innate talent is essentially a Gaussian and the different conditions for their growth turn the talent distribution in society into a power law with a long tail (nurture being more important than nature), the current changed circumstances (of increased life span and wealth) can tilt the balance more in favour of nurture. Many people can afford to grow to their satisfaction in their chosen area without recourse to huge social incentives and support. 

Hence, over time, societal support might diminish. But, doing science from a classical perspective (that is, getting pleasure in satiating a curious mind) is quite enjoyable to the human mind, and there will always be people willing to spend a significant amount of resources to pursue their passion. 

A  scientifically trained mind is always helpful to society. There could be a situation in the future where a substantial number of people hit the right balance of segregating their time into satisfying their passion (indulging in a sort of blue-sky research) and monetising their talent without too much of a compromise on their core activity of improving their talent. Fortunately, in X-ray astronomy, the avenues for revenue generation are relatively easy and straightforward: education (space is always fascinating to a young mind) and industry (many of the techniques used in X-ray detections have good industry applications) are two of the easy options. But, smoothly balancing the requirements of talent growth (needing a relaxed frame of mind in an unhurried setting) and talent monetisation (requiring alertness in managing time, effort and user requirements) could be a difficult task.

\item {\bf How can these insights be used to formulate a strategy to do X-ray astronomy in the future, particularly in India?}\\
$~$\\

I look at this problem with the central idea that new insightful results in X-ray astronomy can be obtained by applying a collective and refined talent.    The question of talent growth also needs to be addressed.

Experiments in X-ray astronomy are not actual science experiments: one cannot alter the objects of experiments in a controlled manner. However, many objects are in the sky at different settings and conditions, providing a fertile sample to select objects to validate any models and assumptions. 
Hence, instead of controlling objects for experimentation, there is a possibility of subtly controlling the observations to falsify models or validate the assumptions. For this approach to succeed, the number of such `experiments' needs to be large, and the questions posed need to be subtle and deep. 

Since space science experiments are expensive, only a few opportunities are currently available, and these opportunities are cornered mainly by projects claiming to make larger and larger telescopes. `Make a big enough telescope, and then anybody can get a breakthrough result' seems the motto. There are several drawbacks in this scheme: it corners all the resources to a few projects; it encourages routine work (of getting results from the current big telescope) and probably influences the lack of talent growth that I am harping about throughout this article; since fund givers (who are also influenced by popular sentiments) may not be specialists in a subject,  this method also `dumbs down' and trivialises the results (`detecting the farthest black hole or the closest planet'). Also, such a `bigger and better' mode closely resembles Ponzi schemes and hence cannot last long.

For a healthy and meaningful growth of X-ray astronomy, an alternate scheme of a large number of thoughtful experiments proposed and executed by a group of very talented people may be necessary. The opportunities to send more niche experiments undoubtedly exist. The world is getting more prosperous, and doing a space experiment has become, relatively speaking, cheaper. It is well within the means of even small universities to be able to finance a good space experiment. The second aspect, a group of talented people coming together to think of some critical observations, might require purposeful efforts to evolve. The current style of doing research where immediate goals are emphasised and encouraged has probably led to the negative vicious cycle of people getting fully occupied with mundane tasks of teaching, writing proposals, and converting the project reports of the students into research papers, resulting in a lower emphasis on creative work thus progressively lowering their individual expertise and ending up doing things of even lower significance.

It is quite possible to evolve a positive virtuous cycle of improving expertise. If a researcher consciously decides to devote a significant amount of time to only core science activities, foregoing the immediate benefits like number of publications, grants, students (and promotion), it is possible to grow the core expertise and produce better results in the long term. It requires the firm conviction that enhanced expertise in a few years will not only be personally very satisfying, but it will also eventually fetch better publications, grants, and students (and even faster promotion). To substantially change the style of functioning of X-ray astronomy, a 
 good fraction (say, 20\%) of researchers need to follow this style. Creative work is inherently satisfying to the human mind, and it can act as a driver for a change such that in future, when the futility of running after short-term goals becomes apparent, such a shift in culture can really happen.
 
 In the Indian context,  opportunities to make and launch X-ray astronomy instruments have increased significantly in recent times. The number of people active in X-ray astronomy has also increased. However, the ability to consistently grow scientific expertise is somewhat lacking, mainly because of the deluge of allied works. Though the number of X-ray astronomers in India has increased, the total number is still sub-critical, thus providing only a limited opportunity to interact with experts in one's own or allied fields. 
 
  From my personal experience, I can conclude that the inherent urge to do creative work is still intact among a large number of X-ray astronomers in India. The bottlenecks to growth are high, as are the demands on their time from day-to-day (mostly trivial) activities. It is, however, possible for a few people to come together and create their own virtual creative bubble and concentrate on some core science activities. If the long-term benefits of such a collaboration are demonstrated at least by a few such groups, it has the potential to accelerate and create the sort of virtuous cycle that I talked about earlier. In the Appendix, I have given words to my flights of fancy on how any individual can grow as a scientist, even in adverse conditions.

\end{itemize}

In conclusion, X-ray astronomy is a fascinating window to the universe, having an unrealised potential to provide insightful information to solve several critical astrophysical problems pertaining to non-thermal astrophysics. It is suggested that a concentrated collective creative work culture will enable a very meaningful growth in X-ray astronomy.

.


\appendix

\section{X-ray astronomer's art of work} 

  X-ray astronomy is a creative art, in particular, a collective creative art. For the healthy growth of X-ray astronomy, it is essential that practising X-ray astronomers proactively concentrate on enhancing their creative abilities, even when the demands on their time for several mundane non-creative, but necessary activities are very high. I am fascinated by the set of tongue-in-cheek advice to practice the creative art of mathematics given by John E. Littlewood, the  British mathematician of the early twentieth century \citep[`Mathematician's art of work',][]{Littlewood1986}. These should perhaps apply to X-ray astronomy, too.

Littlewood gives examples of different creative work styles and suggests that there are no set rules for creative work,  but it would be wise to find out the usual methods and try them out. Then, he goes on to give examples of extreme levels of creativity and posits that    
`between the extremes [of exceptionally brilliant people and dull people], there is the army of people very gifted but short of genius' who possess `an intense conscious curiosity about the subject, with a craving to exercise the mind on it'. 
Littlewood, 
`with a good deal of diffidence', proceeds to give `some practical advice about research and the strategy it calls for', which includes, among other things,
`4 (or  5 at most) hours per day with breaks every hour', 
`either work all out or rest completely', 
`5 and a half days a week' and 
`3 weeks of holiday' in a year. A clear headed distinction is made to define what work is.

 In the context of practising X-ray astronomy in a world full of essential activities that are not necessarily creative work, it would be wise to distinguish between creative work and other necessary work. One easy thumb rule is whatever demands your undivided concentrated effort (writing a paper, critically reading a paper, making a summary of results, writing a code, designing equipment) is creative work, and the rest (attending seminars, attending meetings, giving sermons to students) are not. 
  
  The first task is to snatch from your busy schedule a few hours (3 - 4) daily, at least four to five times a week. It is essential to have this `creative' time contiguously; at least it should not be shorter than an hour. Only planned jobs should be taken up without any multi-tasking. Concentrate on only one topic in a single unit of an hour, without any interruptions. 
  
   Once you have such intense doses of creative work, it will help you deal with other tasks with less effort. Mostly, it is not the non-availability of time that reduces your efficiency; instead, it is the multiple events and issues that compete for your mental space which influence your productivity. After a draining effort to write an inspiring introduction to your forthcoming paper, the `moronic suggestions from your dumb colleagues' in a meeting will not agitate you and snatch away your mental space; you will ignore them and will still be thinking about that clever sentence in the Introduction that you just wrote.
   
   The next effort is to ensure a more significant fraction of creative work in most of your other activities. You need to be quite creative about it; some specific examples are given below.

   \begin{itemize}
   \item {\bf Collaboration}
   \\
   Collaborative work could be a bane to creative work if you are passively trading your expertise to get nth authorship in a 100-author publication,  but it is a great boon if you collaborate with like-minded experts such that after the collaboration, your expertise matures and distinctly sharpens. You can progressively reduce your involvement in `mass-produced' papers and concentrate on meaningful collaborations instead. You may ensure that the person you are collaborating with has good professionalism and decision-making abilities and has sufficient time to invest in the project under consideration. The project needs to be well-defined and should fascinate you. If you are alert when selecting your collaborations, it is likely that over a period of time, you will have a small number of very meaningful collaborations, at least one of which you lead. Though most collaborations can be done remotely, it is a good practice to occasionally have personal meetings with the collaborators, say once in three or six months.
  \item  {\bf Students and post-docs}
  \\
  Many of the extra efforts required to handle students and post-docs can be reduced considerably by a subtle shift in the idea of guiding students. Instead of thinking that it is your responsibility to groom them and `impart' the knowledge you have gained over time, it may be beneficial to treat them as adults and deem them responsible for their own learning. In today's connected world, information and knowledge are freely available, and hence, you may confine your role of guiding and mentoring only to review their work and suggest corrections, if any, periodically; the period of review can progressively increase from a week to a few months once the style of review is established. Instead of pushing them to good research work, try pulling them to your style of work. Initially, try giving them a distinct part of your current research work and make them a co-worker. Instead of motivating them by the lure of seeing their name as the first author of `a' research paper, imparting the sense of fun in doing good research could be more motivational. 
  Students and post-docs should be good collaborators, that is it.
  \item  {\bf Teaching}
   \\
   Teaching, taken as a separate activity, could be a big drain on your time and energy. Here, it would be rewarding to you and the students to introduce your own research activity as a part of the teaching. It may be relatively easy for graduate courses, but there could be ways to cleverly introduce your research as a part of the course, even in undergraduate courses. For example, if your research interest is Gamma-ray Bursts (GRBs) and if you are teaching `Introductory astronomy' for under-grads, the GRB motif can be introduced at various stages like the studies of supernovae, endpoints of stellar evolution, cosmic distance ladder, and cosmology. One of the lectures could be a summary of your own research: it will be an exercise for you to communicate your research to the uninitiated. Also, it can be highly motivating to the students. Again, treating the students as adults and emphasising their need to own up to the responsibility of learning early in their career would be very positive learning.
   \item  {\bf Reviewing}
   \\
   Reviewing and refereeing could drain your time, mainly if you are quite efficient about it. A bit of pragmatic thinking will help. Take up only those works for review if you feel there could be interesting learning. You can be shamelessly ruthless about refusing to referee papers which you think won't add anything to your learning. So long as, on average, the number of papers that you referee equals the number of papers that you submit (not the number of papers you co-author), the law of averages should catch up, and your conscience can be clear.
   \item {\bf Management and administartion}
   \\
Managing a project or administering the running of a department requires a lot of effort: management and administration is a totally different profession compared to doing science. It is very difficult to practice two professions simultaneously: at best, you can have one as a profession and another as a hobby. If you decide to become a professional manager, you have to painfully accept that you no longer wish to be a scientist. Practising science as a hobby and returning to it as a full profession later is extremely difficult. On the other hand, you may do the management as a hobby for a few years and then fully revert back to science. You may have to hire professional managers or secretaries to assist you. It is quite a difficult balancing game, but if you love doing science, there are always ways to stick to it as a profession. The bottom line is it is up to you to make the painful choice, but you can't ride both horses.
  \end{itemize}
   
   The above suggestions are a set of fanciful subjective expansions of the basic tenet that if you have `intense conscious curiosity about the subject, with a craving to exercise the mind on it', as Littlewood puts it, then it is beneficial both to you and the society at large to find ways to satiate that curiosity, even while practising the worldly needs of `using' your expertise for an economically beneficial activity like teaching.

\section*{Acknowledgements}

I thank Gulab Dewangan and N.P.S. Mithun for providing the data for Crab and IC 4329A and 
 helping analyse these data for the sensitivity calculations. 



 

\bibliography{journey}


\end{document}